\documentclass{JHEP3}

\usepackage{graphicx}
\usepackage{fancyvrb}
\usepackage{makeidx}
\usepackage{multirow}
\usepackage{makecell}
\usepackage{url}

\usepackage{latexsym,amssymb,amsmath,graphicx, makeidx}

\newcommand{\be}{\begin{equation}}
\DefineVerbatimEnvironment{code}{Verbatim}{fontsize=\small}
\newcommand{\ee}{\end{equation}}
\newcommand{\bi}{\begin{itemize}}
\newcommand{\ei}{\end{itemize}}
\makeindex

\title{Exploring the Higgs Portal with 10 fb$^{-1}$ at the LHC}

\author{Brian Batell$^a$, Stefania  Gori$^{a,b}$, and Lian-Tao Wang$^{a,c}$\\
        $^a$Enrico Fermi Institute and Department of Physics, University of Chicago, Chicago, IL, 60637\\
	$^b$HEP Division, Argonne National Laboratory, 9700 Cass Ave., Argonne, IL 60439\\
       $^c$Kavli Institute for Cosmological Physics, University of Chicago, Chicago, IL, 60637\\
       E-mail: \email{batell@uchicago.edu, goris@uchicago.edu, liantaow@uchicago.edu}
       }

\abstract{
We consider the impact of new exotic colored and/or charged matter  interacting through the Higgs portal on Standard Model Higgs boson searches at the LHC.
Such Higgs portal couplings can induce shifts in the effective Higgs-gluon-gluon and Higgs-photon-photon couplings, thus modifying the Higgs production and decay patterns. We consider two possible interpretations of the current LHC Higgs searches based on $\sim 5$ fb$^{-1}$  of data at each detector: 1) a Higgs boson in the mass range $(124 -126)$ GeV and 2) a hidden heavy Higgs boson which is underproduced due to the suppression of its gluon fusion production cross section. We first perform a model independent analysis of the allowed sizes of such shifts in light of current LHC data.
As a class of possible candidates for new physics which gives rise to such shifts,  we investigate the effects of new scalar multiplets charged under the Standard Model gauge symmetries.
We determine the scalar parameter space that is allowed by current LHC Higgs searches, and compare with complementary LHC searches that are sensitive to the direct production of  colored scalar states. 
}

\begin{document}
\maketitle
\tableofcontents

\section{Introduction}
\label{sec: introduction}

The minimal Standard Model (SM) posits that electroweak symmetry breaking (EWSB) occurs due to the condensation of an elementary hypercharge $Y=\frac{1}{2}$, $SU(2)_L$ doublet scalar field, leading to the prediction of the SM Higgs boson with well-determined properties~\cite{Higgs:1964ia,Englert:1964et,Higgs:1964pj, Guralnik:1964eu,Weinberg:1967tq} (for a recent review on the SM Higgs boson, see~\cite{Djouadi:2005gi}). 
The SM Higgs boson is the primary target of the LHC, and the  ATLAS and CMS experiments are rapidly closing in on it. The newest results based on $\sim 5$ fb$^{-1}$ of data at each experiment \cite{ATLASHiggs, CMSHiggs} have significantly pushed forward the quest for the Higgs. 

There has been intense investigation on the existence of new physics (NP) at the TeV scale and its possible manifestations. The hierarchy problem in particular suggests that the Higgs must  couple to these new states.
Such couplings can induce significant changes to the Higgs boson phenomenology, affecting both its  production and decay properties. This possibility is especially interesting in light of the recent results from LHC Higgs searches. Both ATLAS and CMS have released the combined SM Higgs limits based on $\sim 5$ fb$^{-1}$ datasets. The SM Higgs boson is now ruled out in the heavy mass range $130 - 600$ GeV. Furthermore, both experiments observe a 2-3$\sigma$ excess in the range 124-126 GeV, which has prompted a large number of studies~\cite{Ibe:2011aa,Arbey:2011ab,Heinemeyer:2011aa,Heinemeyer:2011aa,Li:2011ab,EliasMiro:2011aa,Feng:2011aa,Baer:2011ab,Englert:2011aa,Carena:2011aa,Djouadi:2011aa,Ferreira:2011aa,Guo:2011ab,Moroi:2011aa,Moroi:2011ab,Xing:2011aa,Draper:2011aa,Cheung:2011aa,Hall:2011aa,Kane:2011kj,Kadastik:2011aa,Akula:2011aa,Buchmueller:2011ab,Masina:2011aa,Masina:2011un,Ellwanger:2011aa,Cheung:2011nv,deSandes:2011zs,Baek:2011aa,Arbey:2011aa}.

The size of the dataset is by now large enough to begin probing all regions of SM Higgs masses from $115 - 600$ GeV. As such, LHC Higgs searches are now 
 sensitive to couplings of the Higgs boson to NP states beyond the SM  (BSM).
 For example, the existence of new light states with mass 
$m < m_h / 2$ that couple to the Higgs field opens up new decay channels for the Higgs boson, which is especially relevant for a light Higgs boson, $m_h \leq130$ GeV
\cite{Silveira:1985rk,McDonald:1993ex,Burgess:2000yq,Berger:2002vs,Dermisek:2005ar,Chang:2005ht,Bellazzini:2009xt,Falkowski:2010cm,Kanemura:2010sh,Englert:2011us} (see Ref.~\cite{Chang:2008cw} for a review),
in which case new search strategies may be needed by the collaborations in order to detect the Higgs boson~\cite{Falkowski:2010hi}.
Another possibility is that the Higgs is part of an extended scalar sector, and its coupling is modified due to its mixing with the new scalar particles~\cite{Schabinger:2005ei, O'Connell:2006wi,Barger:2007im,Carena:2011fc,Englert:2011yb}.
In this case, the lightest physical mass eigenstate will have a non standard coupling to SM particles, thus modifying its production rate.  

In this paper, we focus on a scenario in which the Higgs boson is mainly SM-like. Interactions between the Higgs and new exotic particles $S$ lead to modifications of loop-level Higgs production and decay processes, such as $gg \to h$ and $h \to \gamma \gamma$. Additionally a new decay mode $h\to SS$ can be open in the case of a relatively heavy Higgs, leading to small modification of the Higgs total width.
In general there are many possible forms of interactions between the Higgs and the exotics, which depend in detail on the specific quantum numbers of the new exotic states.  
However, there exists a class of interactions which are sufficiently generic and of a universal form so as to deserve special attention: the so-called Higgs portal interactions.
The combination $H^\dagger H$, being a gauge and Lorentz singlet, can be trivially combined with an operator   ${\cal O}_{\rm NP}$, itself a gauge and Lorentz invariant operator built out of exotic new fields. 
The Higgs portal interactions are thus of the generic form
\be
{\cal L} \supset \lambda H^\dagger H {\cal O}_{\rm NP}.
\ee
Furthermore, since $H^\dag H$ is a dimension two operator, the Higgs portal interaction is typically of a low dimension, and may even be renormalizable if new exotic scalar fields are present. 
If the new exotic states 
are  charged under color and/or electromagnetism, integrating them out  generates the operators $ h G^a_{\mu \nu}G^{a \mu \nu}$ and $ h  F_{\mu \nu}F^{ \mu \nu}$. 
As we will demonstrate in detail in this paper, up-to-date LHC results based on the combined 10 fb$^{-1}$ dataset already put interesting constraints on the sizes of such operators.

It is also interesting to consider explicit examples of the NP content. If the NP states are heavy compared to the Higgs boson, 
their interactions can be treated as contact interactions. Additionally, if they are not too heavy, as is often the case for regions with significant modifications to Higgs phenomenology, they can be produced and searched for at the LHC. Combining information from direct searches for exotic states with Higgs searches 
provides complementary probes of the NP. 

In this paper we will focus on the Higgs portal couplings to additional scalars $S$ of the form 
\begin{equation}
\label{eq:portal}
{\cal L} \supset - \lambda (H^\dag H) (S^\dag S),
\end{equation}
and study a large set of possible $SU(3)_C \times SU(2)_L \times U(1)_Y$ representations of $S$.
In principle, additional fermions $F$ and vector bosons $V$ coupling via the Higgs portal can also change the phenomenology of the Higgs boson. However, in order to modify the Higgs phenomenology in a noticeable way with exotic particle loops, the portal couplings cannot be too small in comparison with the SM gauge and top Yukawa couplings. For fermions  and vector bosons,  the most generic  Higgs portal  couplings, $(H^\dag H)(\bar F F)$ and $H^\dagger H V^{\mu \nu} V_{\mu \nu}$, are nonrenormalizable (see however Ref.~\cite{Lebedev:2011iq} for renormalizable models of vector Higgs portals). Hence, in comparison, it appears more natural to expect the marginal operator with scalars in Eq.~(\ref{eq:portal}) to have a sizable coupling. We therefore restrict to new exotic scalars $S$.

This paper is organized as follows. We begin by summarizing the current LHC Higgs searches and our fitting method in Sec.~\ref{sec:Searches}. In Sec.~\ref{sec:EffOps}, we perform a model independent analysis on the allowed sizes of the effective Higgs-gluon-gluon and Higgs-photon-photon operators. The allowed parameter regions for NP scalar particle couplings through the Higgs portal is presented in Sec.~\ref{sec:Scalar}, where we consider both colored and uncolored scalars. Sec.~\ref{sec:Collider} is dedicated to the analysis of the current direct searches of the NP scalars. We shall reserve Sec.~\ref{sec:Conclusion} for our conclusions.

\section{Implications of LHC Higgs searches for new physics}
\label{sec:Searches}

After a very successful year of running in 2011, the LHC has now delivered in total approximately 5 fb$^{-1}$ of integrated luminosity to both the ATLAS and CMS experiments. This is enough data to significantly test the hypothesis of a SM Higgs boson over most of its mass range. The final results of the Higgs searches for the 2011 run have now been released
~\cite{ATLASHiggs, CMSHiggs}. We now give a brief overview of the analysis and results for each experiment in turn.

The ATLAS combination includes the channels
  $h \rightarrow \gamma \gamma$, $h\rightarrow Z Z^* \rightarrow 4 \ell$, 
  $h \rightarrow WW^* \rightarrow 2 \ell 2 \nu$, 
  $h \rightarrow WW^* \rightarrow  \ell  \nu q \bar q'$,
$h \rightarrow ZZ^* \rightarrow 2\ell  2\nu $,  and
$h \rightarrow ZZ^* \rightarrow 2\ell 2 q $, 
which use the full data set of up to 4.9 fb$^{-1}$ of luminosity. 
The combination yields the observed 95$\%$ C.L. exclusion regions, which extend over Higgs masses in the ranges 110.0 - 117.5 GeV, 118.5 - 122.5 GeV and 129 - 539 GeV.
Furthermore, an excess is observed near $m_h \sim 126$ GeV with a local significance of  2.5$\sigma$ (global probability of $30\%$ after accounting for the look-elsewhere effect given the search range of 110-600 GeV). This excess is driven mainly by the $h \rightarrow \gamma \gamma$ and $h\rightarrow Z Z^* \rightarrow 4\ell$ channels. 

The CMS combination includes the channels 
$h \rightarrow \gamma \gamma$,
$h\rightarrow \tau \tau$,
$h\rightarrow bb$,
$h \rightarrow WW^* \rightarrow 2 \ell 2 \nu$, 
$h\rightarrow Z Z^* \rightarrow 4 \ell $,
$h\rightarrow Z Z^* \rightarrow 2 \ell 2 \nu$, 
$h\rightarrow Z Z^* \rightarrow 2 \ell 2 q$, and
$h\rightarrow Z Z^* \rightarrow 2 \ell 2 \tau$, 
all of which have been updated using the full 
dataset of up to 4.8 fb$^{-1}$. CMS is able to exclude the SM Higgs
in the mass range $127-600$ GeV at $95\%$ C.L. 
However, CMS also observes an excess at around $m_h = 124$ GeV 
with a local significance of 3.1$\sigma$ (global significance of 1.5$\sigma$ 
after accounting for the look-elsewhere effect  given the search range of 110-600 GeV), driven by an excess 
of $\gamma \gamma$ events.

From these results we are motivated to consider two distinct hypothetical scenarios: 
\begin{itemize}

\item {\bf Scenario A (Hint of a light Higgs)}: The observed excesses are the first hints of the Higgs boson, with a mass in the (124-126) GeV range. In this case, the excesses observed are, at the moment, certainly consistent with a SM Higgs boson due to the limited statistics. However, the data may suggest the presence of NP which slightly modifies the production and decay patterns, and in particular increases the rate to the diphoton mode by a factor of $\sim 2$. 

\item {\bf Scenario B (Hidden heavy Higgs)}: The Higgs boson is `hidden'.  The mass of the Higgs boson is in the excluded region, but the Higgs particle has so far avoided detection because its production in the most sensitive search channels is suppressed by NP effects compared to the SM predictions. The currently observed excesses at low mass are not due to the Higgs boson, but to a statistical fluctuation.

\end{itemize} 

We will explore both of these scenarios below, first in an effective field theory context followed by specific models of new exotic colored and charged scalar fields coupled via the Higgs portal. 
Before presenting the results, let us first describe the various components of our analysis.


\vspace{5pt}
\noindent {\bf Scenario A (Hint of a light Higgs) }: 
\vspace{5pt}

For Scenario A described above we consider a hypothetical SM Higgs boson in the mass range of $(124-126)$ GeV supplemented with new interactions that affect the Higgs production and decay properties (both effective operators and explicit Higgs portal scalar models). 
We are interested in finding which parameters describing the NP scenarios give an acceptable description of the data, and in particular give a boost to the rate in the diphoton channel. To address this question, for each search channel $i$ we calculate the signal strength parameter,
\begin{equation}
 \frac{\sigma_i^{\rm NP}}{\sigma^{\rm SM}_i},
\end{equation}
where $\sigma_i \equiv \sigma(pp \rightarrow h)\times {\rm Br}(h \rightarrow i) $.
We compare these values with the best-fit values of $\sigma_i / \sigma_i^{\rm SM}$ obtained by the experimental collaborations using a simple $\chi^2$ goodness-of-fit test. We note that 
ATLAS provides such best-fit values for all values of Higgs masses, while CMS provides this information only for $m_h = 119.5$, $124$, $137$, and $144$ GeV. Because of this, as well as the precise values of $m_h$ for which excess occurs, we will consider two separate cases: a) ATLAS data with a $m_h = 126$ GeV hypothesis, and b) CMS data with a $m_h = 124$ GeV hypothesis. We will highlight the differences in ATLAS and CMS results and their corresponding effects on the NP scenarios whenevever possible.

The best-fit values for $\sigma / \sigma^{\rm SM}$ for each experiment and channel are presented in Table~\ref{tab:fit}.  We obtain these values directly from the plots of the best-fit signal strength presented by the ATLAS and CMS collaborations in Refs.~\cite{ATLASHiggs,CMSHiggs}. Notice that in several channels the uncertainties extend to values of negative signal strength, which is not physical. However, in our fits the cross sections (and thus signal strengths) are constrained to be positive as demanded by physics. 
We note that the uncertainties in these measurements are asymmetric, and differ by $\sim 10 - 30  \%$ in some cases. To be conservative in excluding parameters, and for simplicity, we employ a $\chi^2$ goodness-of-fit test using the larger of the errors in each channel, which is shown in Table~\ref{tab:fit}. However, the use of a gaussian likelihood function (as implied by use of the $\chi^2$  statistic) is questionable in the case of asymmetric errors. Therefore, we have also verified that our results do not change significantly when using a two-sided gaussian likelihood, with which both upper and lower error bars can be taken into account.

\begin{table}
{\small
\begin{center}
\renewcommand{\arraystretch}{1.2}
 \begin{tabular}{|c||c|c|c|c|c|}
\hline
   & ~$\gamma \gamma$~ & ~$ZZ$~  & ~$WW$~ & ~$\tau \tau$~ &~$bb$~   \\
 \hline
\hline
ATLAS ( \! \!$m_h \! = \! 126$ GeV \!)  & $1.96\pm 0.86$  & $1.19\pm 1.18$  & $0.17 \pm 0.65$   & $0.14 \pm 1.7$ & $-0.82 \pm 2$    \\
\hline
\! CMS (\! \! $m_h \! = \! 124$ GeV \!)  & $2.1\pm0.62$ & $0.48\pm 1.06$ & $ 0.67 \pm 0.57$  & $0.84 \pm 1.32$  & $1.16 \pm 1.65$      \\
\hline
\hline
\end{tabular}
\end{center}}
\caption{ATLAS ($m_h = 126$ GeV) and CMS($m_h = 124$) best-fit values of $\sigma /\sigma^{\rm SM}$ for the channels used in the fit. 
}
\label{tab:fit}
\end{table}

\vspace{5pt}
\noindent {\bf Scenario B (Hidden heavy Higgs)}:
\vspace{5pt}

Next, for Scenario B described above we consider a hypothetical SM Higgs boson with its mass as a free parameter, again supplemented by NP interactions. 
We are interested in determining parameter regions for which the Higgs boson would not yet have been detected, due to a suppression in the rate in its most sensitive channels caused by NP. To find such regions, we follow a procedure similar to the one outlined in 
Ref.~\cite{Draper:2009fh,Draper:2009au,Weihs:2011wp}. Namely, we compute the {\it observed} $95\%$ C.L. limit on the signal strength parameter in each channel,
\begin{equation}
\mu^{\rm NP} _i \equiv \frac{\sigma^{\rm obs}_i}{ \sigma^{ \rm NP}_i }  =
 \frac{\sigma^{\rm obs}_i }{ \sigma^{\rm SM}_i } \frac{ \sigma^{\rm SM}_i }{ \sigma^{\rm NP}_i } 
 = \mu^{\rm SM}_i \frac{\sigma^{\rm SM}_i }{ \sigma^{\rm NP}_i}.
\label{MuNP}
\end{equation} 
Here $\mu_i^{\rm SM}$ is the factor by which the SM cross section in the channel $i$ (defined as 
$\sigma_i^{\rm SM} \equiv \sigma_i^{\rm SM}(pp \rightarrow h)\times {\rm Br}(h\rightarrow i$))
must be scaled to be excluded at the $95\%$ C.L. .  
The values of $\mu_i^{\rm SM}$ are determined by each collaboration and are extracted from Refs.~\cite{ATLASHiggs,CMSHiggs}. Given these values, Eq.~(\ref{MuNP}) tells us the factor by which the NP cross section in the channel $i$ must be scaled to be excluded at $95\%$ C.L. . 

In order to find the parameter space allowed by the constraints from all channels, we must combine channels. We use the simple prescription of Ref.~\cite{Draper:2009fh,Draper:2009au,Weihs:2011wp} in which 
$\mu^{\rm NP} _i$ parameters for all channels are combined in inverse quadrature to obtain the combined parameter $\mu_{\rm comb}^{\rm NP}$. Then, to be consistent with current constraints at the $95 \%$ C.L., we require  $\mu^{\rm NP}_{\rm comb} > 1$, and this defines the allowed region in our NP parameter space.

\section{Effective operator analysis}
\label{sec:EffOps}

In this section we  derive model-independent constraints on Higgs boson interactions with NP beyond the SM using the up-to-date LHC Higgs boson searches. 
Here we work in the regime in which we can integrate out new exotic particles and represent their effects with contact interactions. We will focus specifically on possible modifications to the processes $gg\rightarrow h$ and $h \rightarrow \gamma \gamma$. Since these processes are loop induced in the SM, they are particularly susceptible to NP effects~\cite{Manohar:2006gz}.  

We parameterize the new effective operators as 
\be
\label{eq:Ops}
{\cal L} \supset   c_G \frac{\alpha_s}{4 \pi v^2} H^\dagger H G^a_{\mu \nu}  G^{a \mu \nu} + c_{\gamma} \frac{\alpha}{4 \pi v^2} H^\dagger H F_{\mu \nu}  F^{\mu \nu}.
\ee
Obviously, it is not always a good approximation to treat these couplings as contact interactions, particularly when the masses of the new exotic particles are comparable to that of the Higgs. In such a situation, the effects of the NP are more model dependent, but we note that there are Higgs boson observables that can signal in a model independent fashion the presence of light degrees of freedom~\cite{Arnesen:2008fb}.
 We do expect that for large regions of parameter space in generic NP models, especially those with a light Higgs boson, the contact operator approximation is valid,  underscoring the general utility of such a model independent analysis. 
 
There are only two free parameters in the effective Lagrangian in Eq. (\ref{eq:Ops}): $c_G$ and $c_\gamma$. In actual NP models, the sizes of the effective operators will depend on several properties of the NP states, i.e., their couplings to the Higgs, their representation under the SM gauge group, number of their species, and their masses. 

The effective operators in the Lagrangian (\ref{eq:Ops}) modify the gluon fusion production cross section as
\be\label{eq:ggcrossop}
\frac{\sigma (gg\rightarrow h)}{\sigma (gg\rightarrow h)_{{\rm{SM}}}}
\sim
\frac{\Gamma (h\rightarrow gg)}{\Gamma (h\rightarrow gg)_{{\rm{SM}}}}
=
\left|1+\frac{4 c_G}{\sum_f A_{1/2}(\tau_f)}\right|^2\,,
\ee
as well as the partial Higgs decay width to a pair of photons, 
\be\label{eq:gammagammaBRop}
\frac{\Gamma (h\rightarrow \gamma\gamma)}{\Gamma (h\rightarrow\gamma\gamma)_{{\rm{SM}}}}
=
\left| 1-\frac{2c_\gamma}{A_1(\tau_W)-\sum_f N_f Q_f^2 A_{1/2}(\tau_f)}\right|^2\,.
\ee
In Eqs.~(\ref{eq:ggcrossop},\ref{eq:gammagammaBRop}) above, 
 $\tau_i \equiv m_h^2/4 m_i^2$, $A_{1/2}$ ($A_1$) a fermion (vector boson) loop function defined in the Appendix, 
and  $N_f$ and $Q_f$ the number of colors and the electric charge of the fermion $f$ of the SM.
We now consider in turn the two distinct interpretations of the LHC Higgs searches, as described in Section~\ref{sec:Searches}.

\vspace{5pt}
\noindent {\bf Scenario A (Hint of a light Higgs)}:
\vspace{5pt}

Consider first the Scenario A, in which we hypothesize the existence of a Higgs boson in the mass range (124 - 126) GeV. The observed deviations are consistent with a SM Higgs boson, but the best-fit values of the rate in the diphoton channel are roughly a factor of 2 larger than the one predicted by the SM. It is interesting to ask what values of $c_G$ and $c_\gamma$ lead to this boosted rate. Moreover, it is of general interest to determine the constraints on the sizes of the operators $c_G$ and $c_\gamma$ in Eq.~(\ref{eq:Ops}), assuming a Higgs boson is present in this mass range. Such constraints can then be easily translated to any NP model. 

\begin{figure}[h!]
\centerline{
\includegraphics[width=.48\textwidth]{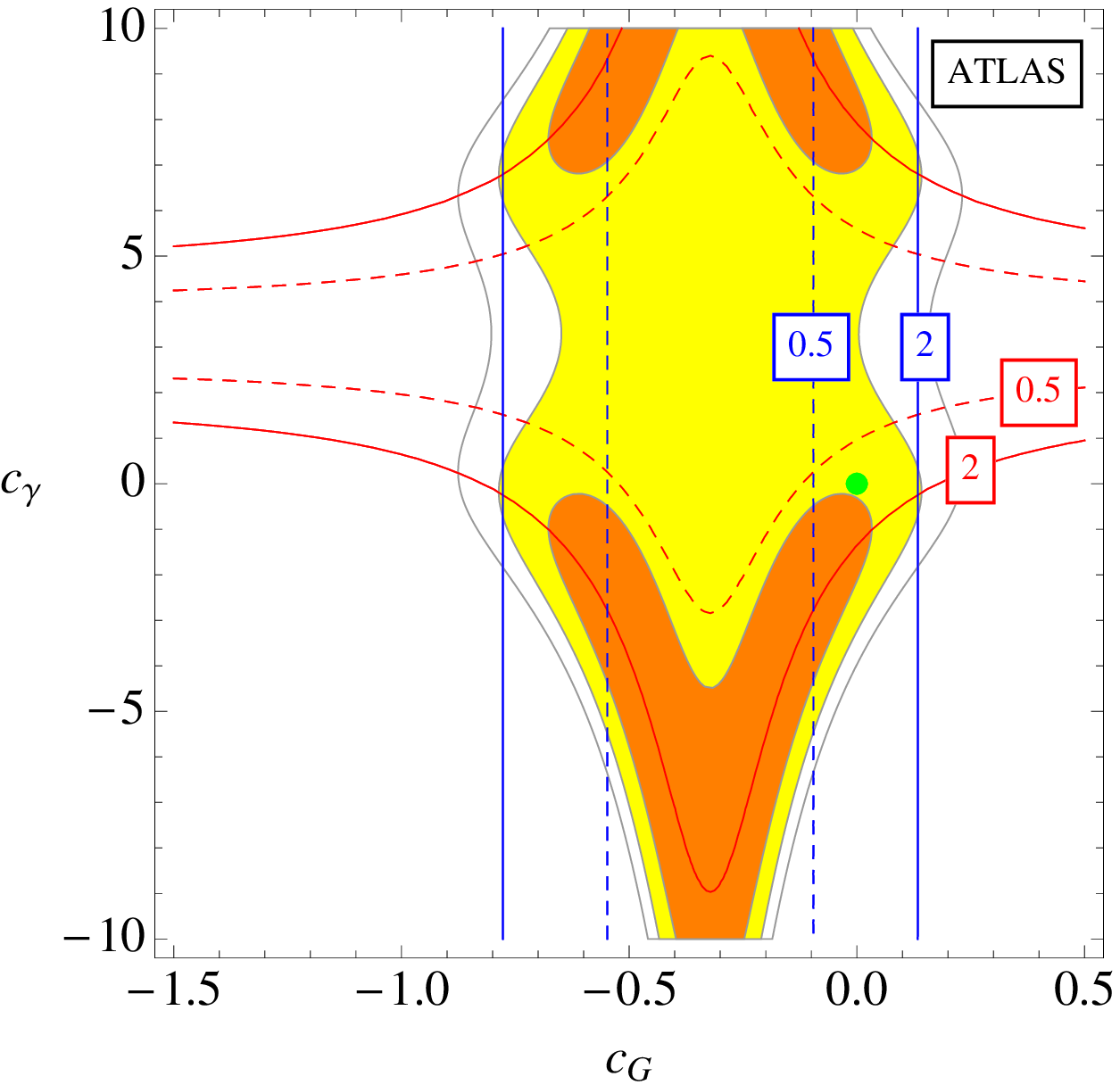} \quad
\includegraphics[width=.48\textwidth]{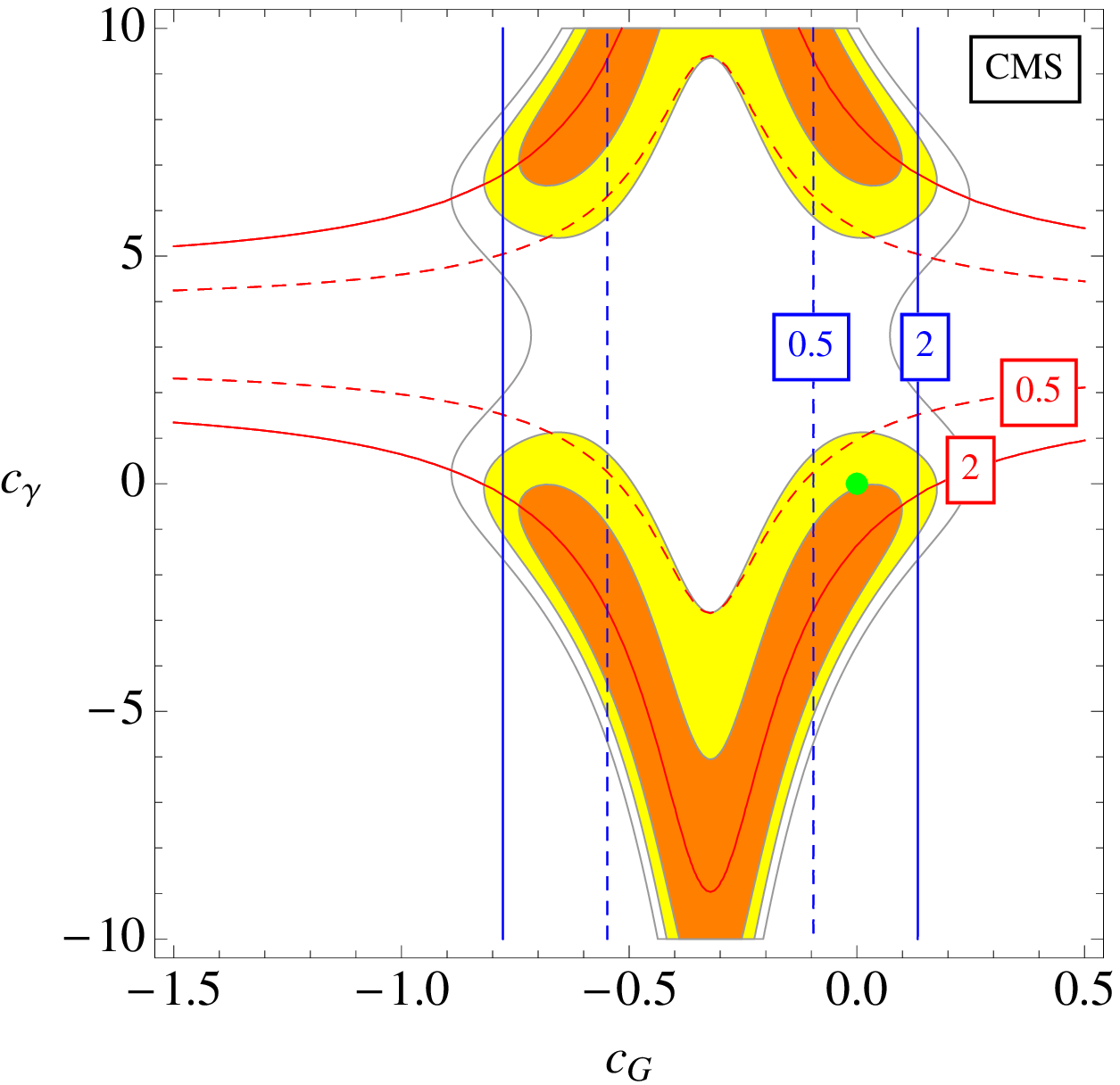} }
\caption{ Constraints on the coefficients of the effective operators $c_G$ and $c_\gamma$. In the left panel we show constraints from the ATLAS data for a 126 GeV Higgs boson. In the right panel we show constraints from the CMS data for a 124 GeV Higgs boson. 
In each panel, we show the confidence levels of 1$\sigma$ (orange),  2$\sigma$ (yellow), 3$\sigma$ (solid boundary). We also show contours of constant gluon fusion Higgs production cross section normalized to the SM value  (blue)  and the normalized rate of the $\gamma \gamma$ channel (red). Solid lines show twice the SM rate, while dashed lines show one half the SM rate. The green dot indicates the SM ($c_G = c_\gamma = 0$)}
\label{fig:Ops_1}
\end{figure}

Following the analysis steps outlined in Section~\ref{sec:Searches}, we find the allowed parameter regions in the $c_G  - c_\gamma$ plane. We consider two cases: 1) a fit to the ATLAS data with a 126 GeV Higgs boson, and 2) a fit to the CMS data with a 124 GeV Higgs boson. The results are shown in Fig.~\ref{fig:Ops_1}
\footnote{In the range of $c_G$ and $c_\gamma$ we consider here, the modification of the Higgs total width is negligible.}. 

We now make several remarks regarding the results presented in Fig.~\ref{fig:Ops_1}. 
The general shape of the regions is dictated by two competing factors: First, the best fit regions tend to lie along curves in which the diphoton rate is $\sim$2 times its SM value. We have overlaid contours showing the enhancement (or suppression) of the diphoton rate in red. Secondly, values in which the gluon fusion cross section is too large compared to the SM value lead to tension with the rate observed in the $ZZ$ and $WW$ channels. Comparing the two experiments, we see that ATLAS generally allows a larger suppression of the gluon fusion production cross section than CMS due to the smaller best-fit rate in the $WW$ channel (see Table~\ref{tab:fit}). We show contours of the enhancement (or suppression) of the gluon fusion rate in blue. We observe that in general, NP contributions to gluon fusion are quite constrained, with $-0.7 \lesssim c_G \lesssim 0.1$ (for a $1\sigma$ fit), while $c_\gamma$ is far less constrained.

\vspace{5pt}
\noindent {\bf Scenario B (Hidden heavy Higgs)}:
\vspace{5pt}

We next consider the hypothesis that the Higgs boson is heavy but hidden from the LHC searches thus far due to a NP suppression of its rate in the most sensitive channels\footnote{A heavy Higgs boson leads to a conflict with precision electroweak data, but there are a number of way in which NP may help to relax such a tension~\cite{Peskin:2001rw}. }. We focus on the region $m_h > 130$ GeV, in which case the main relevant channels are $WW$ and $ZZ$. For such heavy Higgs bosons, only the gluon fusion production cross section can be modified by the operators in Eq.~(\ref{eq:Ops}), and therefore the LHC Higgs searches can probe the coefficient $c_G$. 
\begin{figure}[h!]
\centerline{
\includegraphics[width=.6\textwidth]{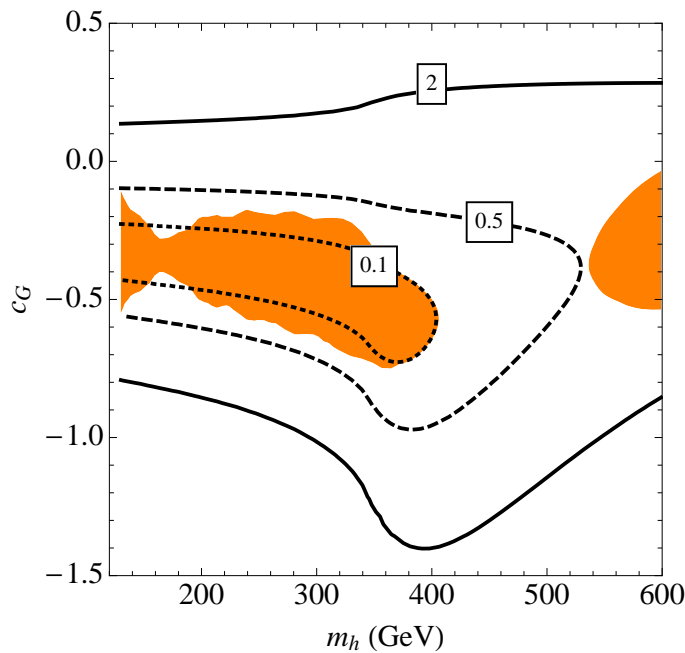} }
\caption{  Values of the coefficient $c_G$ as a function of Higgs mass for which the Higgs boson production cross section is sufficiently suppressed for it to be hidden from current LHC  searches (Orange).  We also show the contours of constant cross section in the gluon fusion channel relative to its SM value: twice (solid), one half (dashed), and one tenth (dotted) the SM values.
}
\label{fig:Ops_11}
\end{figure}

In Fig.~\ref{fig:Ops_11} we  present regions in the $m_h - c_G$ plane for which the Higgs boson would not have yet been detected. To obtain these allowed regions, we follow the analysis steps described in Section~\ref{sec:Searches}. We see that consistency with the null searches requires a sizable NP contribution to the gluon fusion channel, with preferred values of the operator coefficient $-0.7 \lesssim c_G \lesssim -0.1$ for $m_h \lesssim 400$ GeV, while a slightly  larger range  of $c_G$ values are allowed for $m_h\geq 520$ GeV due to the weaker constraints on the SM Higgs production cross section at high masses. Such negative values of $c_G$ lead to the destructive interference with the dominant SM top quark loop contribution in the gluon gluon Higgs production cross section (see Eq.~(\ref{eq:ggcrossop})). Finally, for Higgs masses between $400\,{\rm{GeV}}\lesssim m_h\lesssim 520$ GeV it is not possible to hide a Higgs with the effective operators presented in Eq.(\ref{eq:Ops}). There are two reasons for this: 1) for Higgs bosons heavier than twice the top mass the $gg\rightarrow h$ amplitude develops a large imaginary part that cannot be cancelled with (\ref{eq:Ops}), and 2) the ATLAS constraint in the $ZZ$ channel is particularly stringent in this mass range.

We also overlay contours of constant cross section of the gluon fusion channel relative to its value in the SM. We observe that a sizable suppression 
$\sigma_{gg\rightarrow h}  \lesssim 0.1 \sigma_{gg\rightarrow h}^{\rm SM} $ is required for $m_h \lesssim 400$ GeV, while a smaller suppression is needed at $m_h\geq 520$ GeV, again due to the weaker limits. We conclude that in the most part of parameter space it is possible to rescue a heavy Higgs boson from  strong constraints imposed by LHC searches through NP affecting its production.

\section{Higgs portal to exotic scalars}
\label{sec:Scalar}

We now consider concrete UV completions of the higher dimensional operators we discussed in the previous section.
In particular, we will investigate a very general scenario in which new 
exotic scalar fields in a variety of $SU(3)_C \times SU(2)_L \times U(1)_Y$ representations
interact through the Higgs portal. 
As explained in Sec.~\ref{sec:Searches}, our goal is twofold. 
We first show  which ranges of masses and couplings of the additional scalars best fit a Higgs boson with a mass in the range $(124-126)$ GeV, as recently hinted by the ATLAS and CMS collaborations~\cite{ATLASHiggs, CMSHiggs}. Secondly, we analyze to what extent a relatively heavy Higgs ($m_h\geq 130$ GeV) is still compatible with the current LHC constraints thanks to the presence of the additional scalars, which can lead to the
suppression of the gluon-gluon fusion production cross section as well as the possible decay $h\to SS$ that suppresses the branching ratios of the Higgs to SM particles. Once the preferred parameter space is found,  in Section~\ref{sec:Collider} we will confront it with the parameter space still allowed by the LHC direct searches of these colored scalars. 
 We note that the suppression of the gluon-gluon fusion cross section can occur in various beyond the SM scenarios, such as SUSY~\cite{Djouadi:1998az,Carena:2002qg,Dermisek:2007fi} and composite Higgs models~\cite{Falkowski:2007hz,Low:2010mr}.
Recent studies considering the possibility of suppressing the gluon fusion cross section with colored scalars include~\cite{Bai:2011aa,Dobrescu:2011aa}. 

The Lagrangian of the scalar multiplet contains a kinetic term with the appropriate covariant derivative, as well as a scalar potential containing a bare mass term and possible cubic and quartic couplings. In particular, for all models a Higgs portal coupling is present:
\be
{\cal L} \supset  -\lambda  |S|^2  H^\dag H\, ,
\label{HP}
\ee
in which we have omitted the $SU(3)_C$ and $SU(2)_L$ indices and color contractions are implicit. After electroweak symmetry breaking, the physical squared mass $m_S^2$ of the scalar field is the sum of the bare mass and a contribution from the Higgs portal coupling in Eq.~(\ref{HP}) \footnote{The multiplets charged under $SU(2)_L$ will in general obtain a mass splitting due to radiative effects on the order of hundreds of MeV. Such a small mass splitting will not affect our results.}.
All modifications to Higgs physics are thus described 
by only three parameters: $m_h, m_S, \lambda$.

There are of course various conditions that the couplings in the scalar potential must satisfy, such as that $SU(3)_C$ is not broken and that there are no runaway directions at large field values. These conditions are quite model dependent, since the terms allowed in the potential depend on the specific representations. However, some general comments are in order. First,  the quartic terms involving only the scalar field, such as $\lambda_S (|S|^2)^2$, must have positive couplings $\lambda_S>0$  so that the potential is bounded from below.  
Furthermore, since we will be considering both positive and negative values of the Higgs portal coupling $\lambda$, a runaway direction may develop along a direction of nonzero $H$ and $S$ field values, unless $\lambda$ is greater than some minimum (negative) value~\cite{Bai:2011aa}.  Finally in order to avoid the presence of color or charge breaking vacuua, we require $m_S^2 > 0$. We will then restrict to scalar fields that do not modify electroweak symmetry breaking (i.e. we do not consider electroweak multiplets that obtain a vacuum expectation value). These conditions can always be satisfied for the parameter regions we consider. One may additionally desire to have a viable theory up to some scale larger than the TeV scale, in which case the absolute value of $\lambda$ will be bounded from above (see~\cite{Bai:2011aa} for details).

As to the specific $SU(3)_C\times SU(2)_L \times U(1)_Y$ representations we consider, our aim with this study is to obtain a general overview of the possible modifications of the Higgs phenomenology and the allowed scalar parameter space, as well as the complementary signals of scalar pair production at colliders. 
There are, however, some basic restrictions on the possible representations. 
For example, heavy colored and/or charged particles with TeV-scale masses that are cosmologically stable are strongly disfavored by heavy element searches~\cite{Smith:1982qu}. Furthermore, even if such colored or charged states are only collider stable, stringent lower bounds on their masses exist from LHC searches for stable charged/and or hadronizing particles~\cite{HSCPs}. Colored/charged states that can promptly decay will be much less constrained by direct collider searches. For this reason we choose to study representations in which renormalizable interactions mediating the decay of the new colored/charged states are allowed by the SM gauge symmetries. This restricts the $SU(3)_C\times SU(2)_L \times U(1)_Y$ quantum numbers of the colored/charged scalars. However, we note that this is not a necessary condition: it is possible that nonrenormalizable operators mediate prompt decays of the new scalars. Therefore, our choice is mainly motivated by simplicity. 

Even with this simplifying requirement, we are led to consider a variety of representations with different quantum numbers. This will allow us to obtain a representative overview of the possible modifications of the Higgs phenomenology, thanks to the presence of new colored and EW charged scalars with a mass of a few hundred GeV. The scalar representations we consider are presented in Table~\ref{tab:reps} (see Ref.~\cite{DelNobile:2009st} for a recent comprehensive study of the scalars with nonzero hypercharge).
\begin{table}[h!]
\begin{center}
\renewcommand{\arraystretch}{1.3}
 \begin{tabular}{| c | c | c |}
\hline
\hline
Model
    &  Couplings   & Signatures    \\
 \hline
\hline
  $ ( {\bf 1}, {\bf 1} ,1 )$   
&  $LL$  
&  $(\ell^- \ell^+  )(\ell^- \ell^+  ), \ell^- \ell^+  +\!\not\!\! E_T$  \\ \hline
     $ ( {\bf 1}, {\bf 1} ,2 )$   
&  $e_R e_R$   
&  $(\ell^- \ell^+  )(\ell^- \ell^+  )$ \\ \hline
 \multirow{3}{*}{   ~ $ ( {\bf 1}, {\bf 2} ,\frac{1}{2} )$~~  } 
& $\bar u_R Q $  
& $(jj)(jj), (t j)(\bar t j), (b j)(\bar b j), (t \bar b) (\bar t b), (t \bar t)(t \bar t)$
\\ 
\cline{2-3}     
&  $\bar Q d_R$ 
& $(jj)(jj), (t j)(\bar t j), (b j)(\bar b j), (t \bar b) (\bar t b), (b \bar b)(b \bar b)$ \\ 
 \cline{2-3}   
&  $\bar L e_R$ 
& $(\ell^- \ell^+  )(\ell^- \ell^+  ), \ell^- \ell^+  +\!\not\!\! E_T$  
\\ \hline
  $ ( {\bf 1}, {\bf 3} ,1 )$   
&  $LL$  
& $(\ell^- \ell^+  )(\ell^- \ell^+  ), \ell^- \ell^+  +\!\not\!\! E_T$ \\ \hline
 \multirow{3}{*}{   ~ $ ( {\bf 3}, {\bf 1} ,-\frac{1}{3} )$~~  } 
& $Q Q$, $u_R d_R $  
& $(jj)(jj)$, $(tj)(\bar t j)$, $(bj)(\bar b j)$, $(t b)(\bar t \,\bar b)$ 
\\ 
\cline{2-3}    
&  $\bar Q \bar L$ 
& $ (\ell^- j)(\ell^+ j), (\ell^- t)(\ell^+ \bar t),   2 j+ \! \not \!\! E_T,  b \bar b + \! \not \!\! E_T $ \\  \cline{2-3}   
&  $\bar u_R \bar e_R$ 
& $(\ell^- j  )(\ell^+ j), (\ell^- t)(\ell^+ \bar t)$  
\\ \hline
  $ ( {\bf 3}, {\bf 1} ,\frac{2}{3} )$   
&   $d_R d_R$  
& $   (jj)(jj),(bj)(\bar b j) $ \\ \hline
 \multirow{2}{*}{ $ ( {\bf 3}, {\bf 1} ,-\frac{4}{3} )$  } 
&  $u_R u_R$  
& $(jj)(jj), (tj)(\bar t j) $
\\ 
\cline{2-3} 
&  $\bar d_R \bar e_R$  
&  $(\ell^- j)(\ell^+  j), (\ell^-  b)(\ell^+ \bar b) $  
 \\ \hline
 \multirow{2}{*}{ $ ( {\bf 6}, {\bf 1} , \frac{1}{3} )$}   
& $ \bar Q \bar Q$ 
&  $(jj)(jj)$, $(tj)(\bar t j)$, $(bj)(\bar b j)$
 \\ 
\cline{2-3}  
& $\bar u_R  \bar d_R $  
&  $(jj)(jj)$, $(tj)(\bar t j)$, $(bj)(\bar b j)$, $(t b)(\bar t \,\bar b)$  
\\ \hline
  $ ( {\bf 6}, {\bf 1} , -\frac{2}{3} )$   
& $ \bar d_R  \bar d_R $    
& $(j j)(j j), (b j)(\bar b j), (bb)(\bar b \bar b) $
  \\ \hline
   $ ( {\bf 6}, {\bf 1} , \frac{4}{3} )$  
 & $ \bar u_R  \bar u_R $ 
& $(j j)(j j), (t j)(\bar t j), (t t)(\bar t \bar t) $   
 \\ \hline
   $ ( {\bf 8}, {\bf 1} , 0 )$   & loop decay & $(jj)(jj)$ \\ \hline
  $ ( {\bf 3}, {\bf 2} ,\frac{1}{6} )$   &   $\bar d_R L $   
& $ (\ell^- j)(\ell^+ j), (\ell^- \bar b)(\ell^+  b), 2j+\!\not \!\!E_T , b \bar b+\!\not \!\!E_T $
  \\ \hline
  \multirow{2}{*}{ $ ( {\bf 3}, {\bf 2} ,\frac{7}{6} )$ }  
&   $\bar u_R L $  
&  $(\ell^- j)(\ell^+ j), (\ell^- \bar t )(\ell^+  t), 2j+\!\not \!\!E_T , t \bar t+\!\not \!\!E_T$ 
\\ \cline{2-3}
&  $ \bar Q e_R$ 
& $ (\ell^- j)(\ell^+ j), (\ell^- \bar t )(\ell^+  t),(\ell^- \bar b )(\ell^+  b)$
\\ \hline
 \multirow{2}{*}{ $ ( {\bf 8}, {\bf 2} ,\frac{1}{2} )$ } 
&   $\bar u_R Q $  
& $(jj)(jj), (t j)(\bar t j), (b j)(\bar b j), (t \bar b) (\bar t b), (t \bar t)(t \bar t)$
  \\ \cline{2-3} 
 &  $ \bar Q d_R $ 
& $(jj)(jj), (t j)(\bar t j), (b j)(\bar b j), (t \bar b) (\bar t b), (b \bar b)(b \bar b)$
\\ \hline
\multirow{2}{*}{$ ( {\bf 3}, {\bf 3} ,-\frac{1}{3} )$}   
&   $ QQ $     
 & $(j j)(j j), (t j )(\bar t j), (b j)(\bar b j)$
\\ \cline{2-3}
 & $\bar Q \bar L $
& $(\ell^- j)( \ell^+ j), (\ell^- t)(\ell^+ \bar t),  (\ell^- b)(\ell^+ \bar b), jj+\!\not\!\! E_T, t \bar t+\!\not\!\! E_T, b \bar b+\!\not\!\! E_T$
\\ \hline
 $ ( {\bf 6}, {\bf 3} ,\frac{1}{3} )$   &   $ \bar Q \bar Q $    &  
$(j j)(j j), (t j )(\bar t j), (b j)(\bar b j), (tt)(\bar t \bar t), (bb )(\bar b \bar b), (tb)(\bar t \bar b)
$ \\ \hline
   $ ( {\bf 8}, {\bf 3} ,0 )$   &   loop decay &  
$(W^+ j)(W^- j), (\gamma j)(\gamma j), (Z j )(Z j), (\gamma j)(Z j)$
\\  \hline
\hline
  \end{tabular}
\end{center}
\caption{$SU(3)_C\times SU(2)_L \times U(1)_Y$ scalar multiplets. These scalars have renormalizable interactions mediating their decay and do not contribute to electroweak symmetry breaking. The scalar quantum numbers, couplings to SM matter, and the possible signatures of scalar pair production at colliders are displayed. }
\label{tab:reps}
\end{table}

The three main effects of the colored and electrically charged scalars relevant for the Higgs phenomenology are the modification of the production cross section through gluon-gluon fusion, the modification of the branching ratio into two photons, and the possible decay $h\to SS$ that can increase the Higgs total width. In particular, for the case $m_h<2 m_S$ and not too large values of $\lambda$, the modification to the total width is negligible, and at LO we obtain
\be\label{eq:ggcross}
\frac{\sigma (gg\rightarrow h)}{\sigma (gg\rightarrow h)_{{\rm{SM}}}}\sim\frac{\Gamma (h\rightarrow gg)}{\Gamma (h\rightarrow gg)_{{\rm{SM}}}}=\left|1+Kc\frac{\lambda\, C(r)A_0(\tau_S)\frac{v^2}{m_S^2}}{\sum_f A_{1/2}(\tau_f)}\right|^2,
\ee
in which $\tau_i=m_h^2/4 m_i^2$, $C(r)$ is the Casimir of the $SU(3)_C$ representation and $A_0$ ($A_{1/2}$) a scalar (fermion) loop function given in the Appendix, $c = 1 (1/2)$ for a complex (real) scalar and $K=1,2,3$ in the case of a weak singlet, doublet, triplet.

Similarly the partial width of the Higgs into two photons is modified by
\be\label{eq:gammagammaBR}
\frac{\Gamma (h\rightarrow \gamma\gamma)}{\Gamma (h\rightarrow \gamma\gamma)_{{\rm{SM}}}}=\left|1-c\sum_i \frac{\lambda\, d(r)Q_{S_i}^2A_0(\tau_S)\frac{v^2}{2m_S^2}}{A_1(\tau_W)-\sum_f N_f Q_f^2 A_{1/2}(\tau_f)}\right|^2,
\ee
where the sum is performed over the charged components of the $SU(2)_L$ multiplet~\footnote{Note that the formulas we are giving in this section are valid only in the hypothesis that all the components of the $SU(2)_L$ multiplet are approximately degenerate in mass.}, $d(r)$ is the dimension of the $SU(3)_C$ representation, $N_f$ and $Q_f$ the number of colors and the electric charge of the fermion $f$ of the SM and $A_1$ the gauge boson loop function given in the Appendix. As well known, in the SM the main contribution comes from the $W$ boson loop, so the quantity in the denominator is positive. From Eqs.~(\ref{eq:ggcross}),~(\ref{eq:gammagammaBR}) it is interesting to note that a negative NP effect in the partial width to two photons ($\lambda>0$) corresponds to a positive NP effect in gluon-gluon fusion and vice-versa.

To estimate the quantitative effects on Higgs boson searches, we use the ratios in Eqs.~(\ref{eq:ggcross}), (\ref{eq:gammagammaBR}) to scale the SM gluon fusion cross section and the Higgs to diphoton branching ratio, where the SM values for all cross sections and branchings ratios are taken from the LHC Higgs cross section working group~\cite{HiggsXsection}. 
These ratios are computed at the LO. There have been a few works examining the higher order effects of new colored scalar particles on the gluon fusion process~\cite{Bonciani:2007ex,Boughezal:2010ry}. 
In particular, Ref.~\cite{Bonciani:2007ex} computed the NLO corrections to the gluon fusion Higgs production cross section for a general scalar multiplet, and studied numerically some specific example models. Using their results for the  
$({\bf 8},{\bf 2},\frac{1}{2})$ scalar, one may estimate that the size of NLO corrections could differ from those in the SM by at most 10-20$\%$. Corrections of this size are numerically important, but will not qualitatively alter our results and conclusions at this early stage of limited statistical precision. However as the measurement of the Higgs boson production rates improves with more data, it will be important to take such higher-order effects into account.

Finally, as we already mentioned, in the case of a heavy Higgs, $m_h>2 m_S$, the $h\to SS^*$ decay mode is open, with a partial decay width given by
\begin{equation}
\Gamma(h\to SS^*)=\frac{d(r) \lambda^2 v^2}{16 \pi m_h} \sqrt{1 - 4 \frac{m_S^2}{m_h^2}}.
\end{equation}
 For Scenario B we must take into account this channel, in addition to the modifications of the gluon fusion production cross section presented above.

We now explore the scalar parameter space for the two scenarios described in Section~\ref{sec:Searches}. Namely, we consider first the hypothesis of a light Higgs boson in the mass range $124-126$ GeV, followed by the possibility of a hidden heavy Higgs boson. To illustrate the main features of the constraints on the scalar parameter space, we discuss in detail representative examples in this section: two colored scalar multiplets,  $({\bf 3}, {\bf 1}, -\frac{4}{3})$ and $({\bf 8},{\bf 2},\frac{1}{2})$, as well as one color neutral multiplet $({\bf 1},{\bf 1}, {2})$ for Scenario A and two colored scalar multiplets, $({\bf 3}, {\bf 2}, \frac{1}{6})$ and $({\bf 8},{\bf 2},\frac{1}{2})$ for Scenario B.

\vspace{5pt}
\noindent {\bf Scenario A (Hint of a light Higgs)}:
\vspace{5pt}

\begin{figure}[h!]
\centerline{
\includegraphics[width=.48\textwidth]{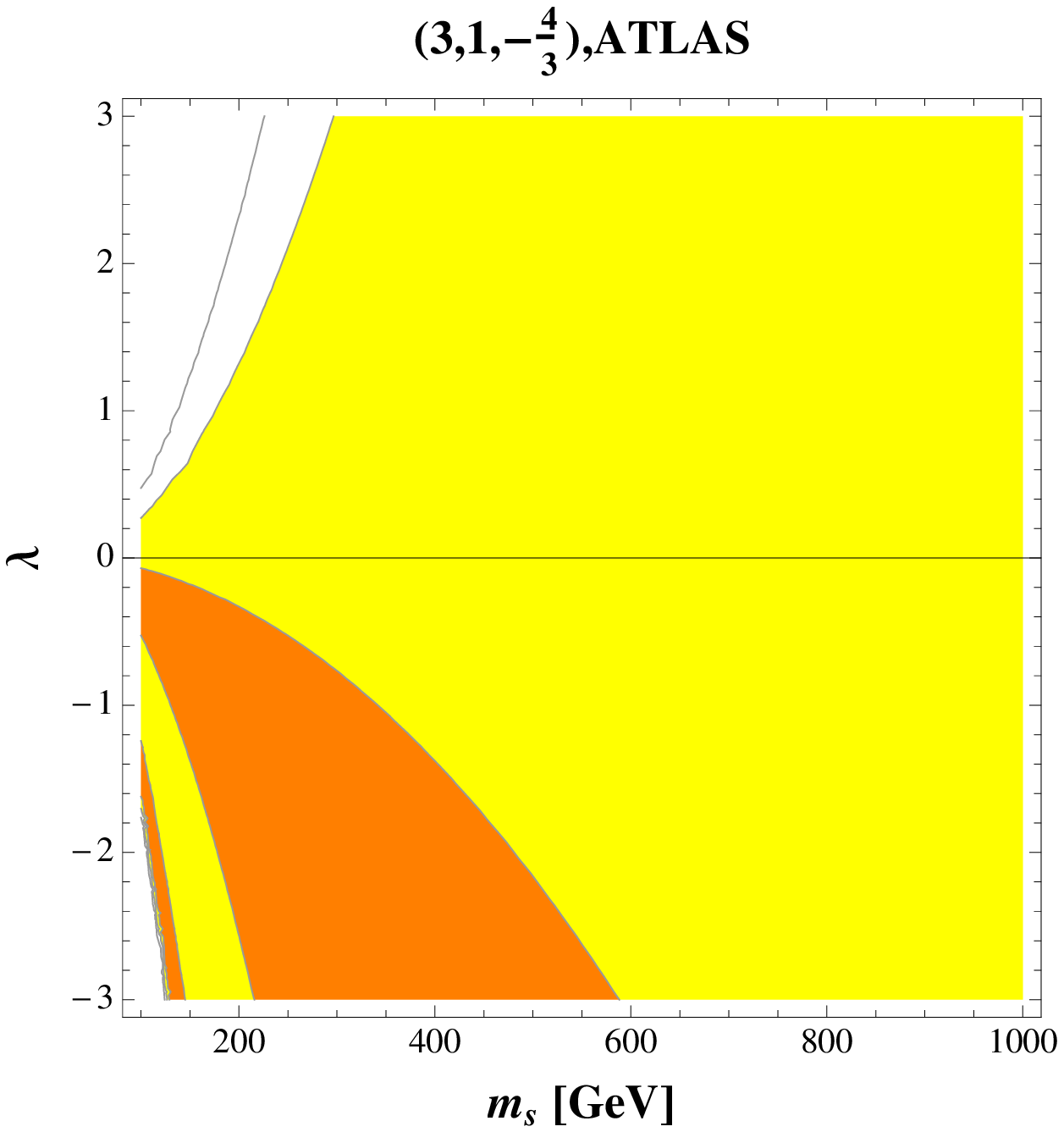}
\includegraphics[width=.48\textwidth]{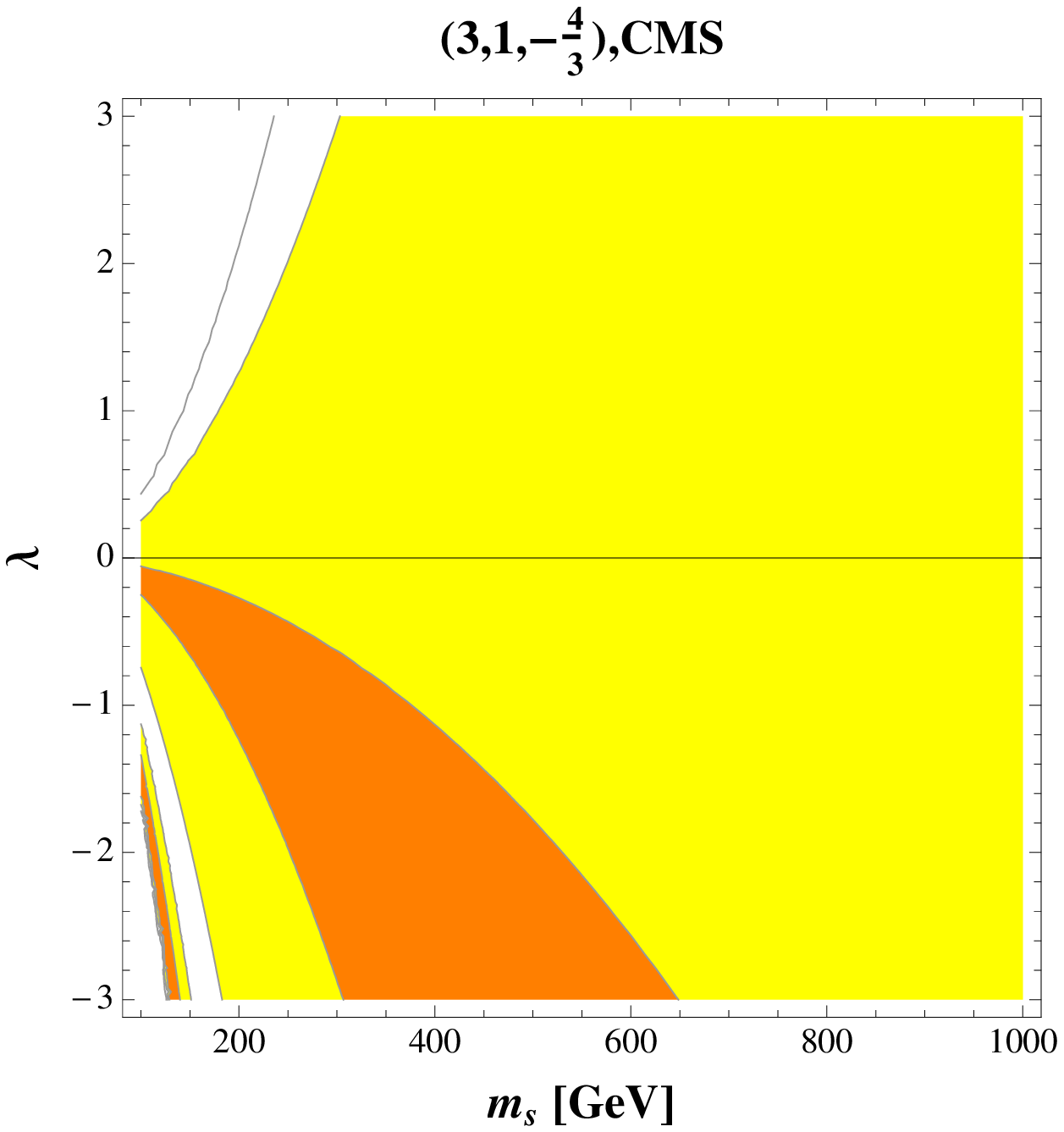} } 
\centerline{
\includegraphics[width=.48\textwidth]{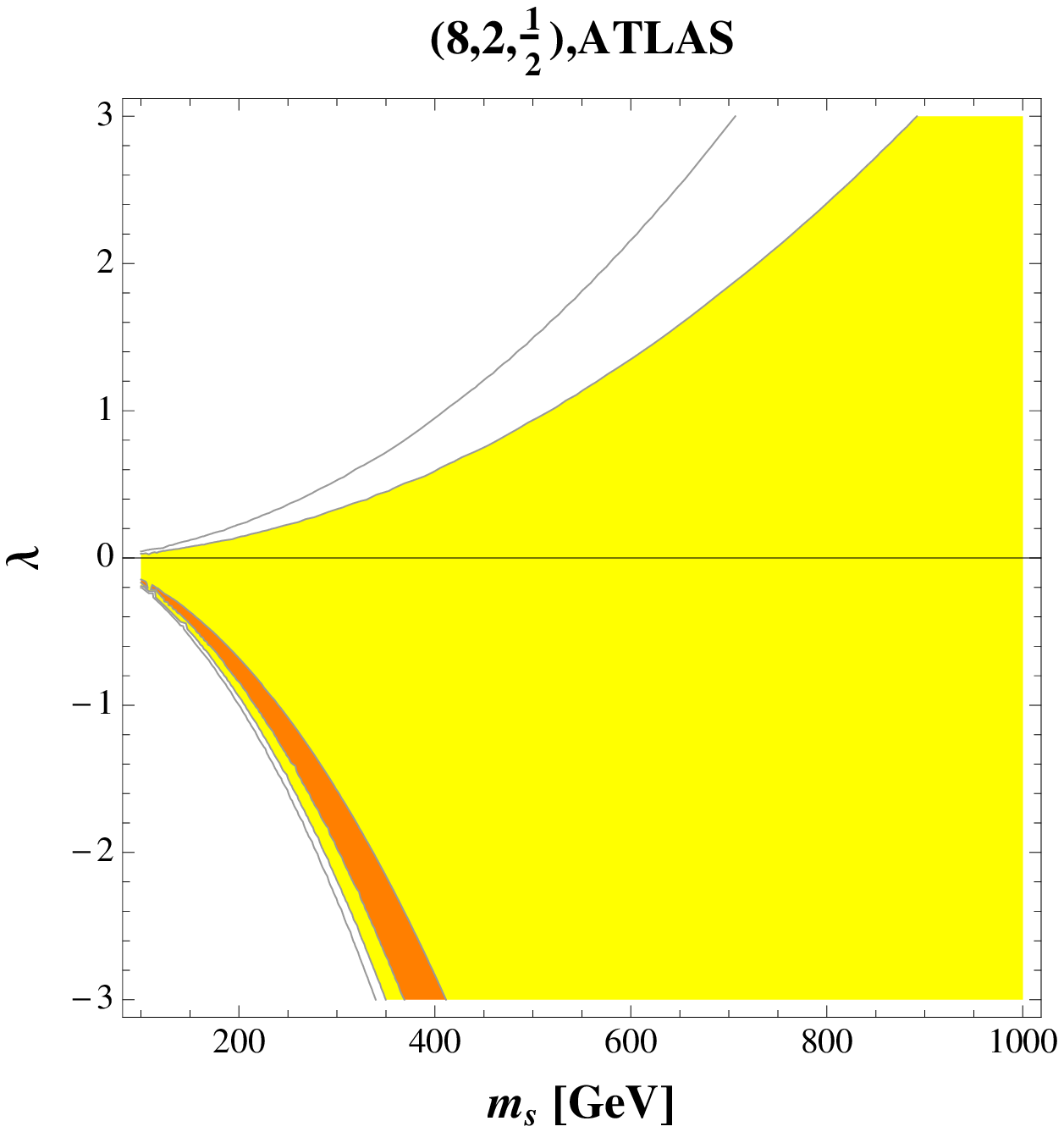}
\includegraphics[width=.48\textwidth]{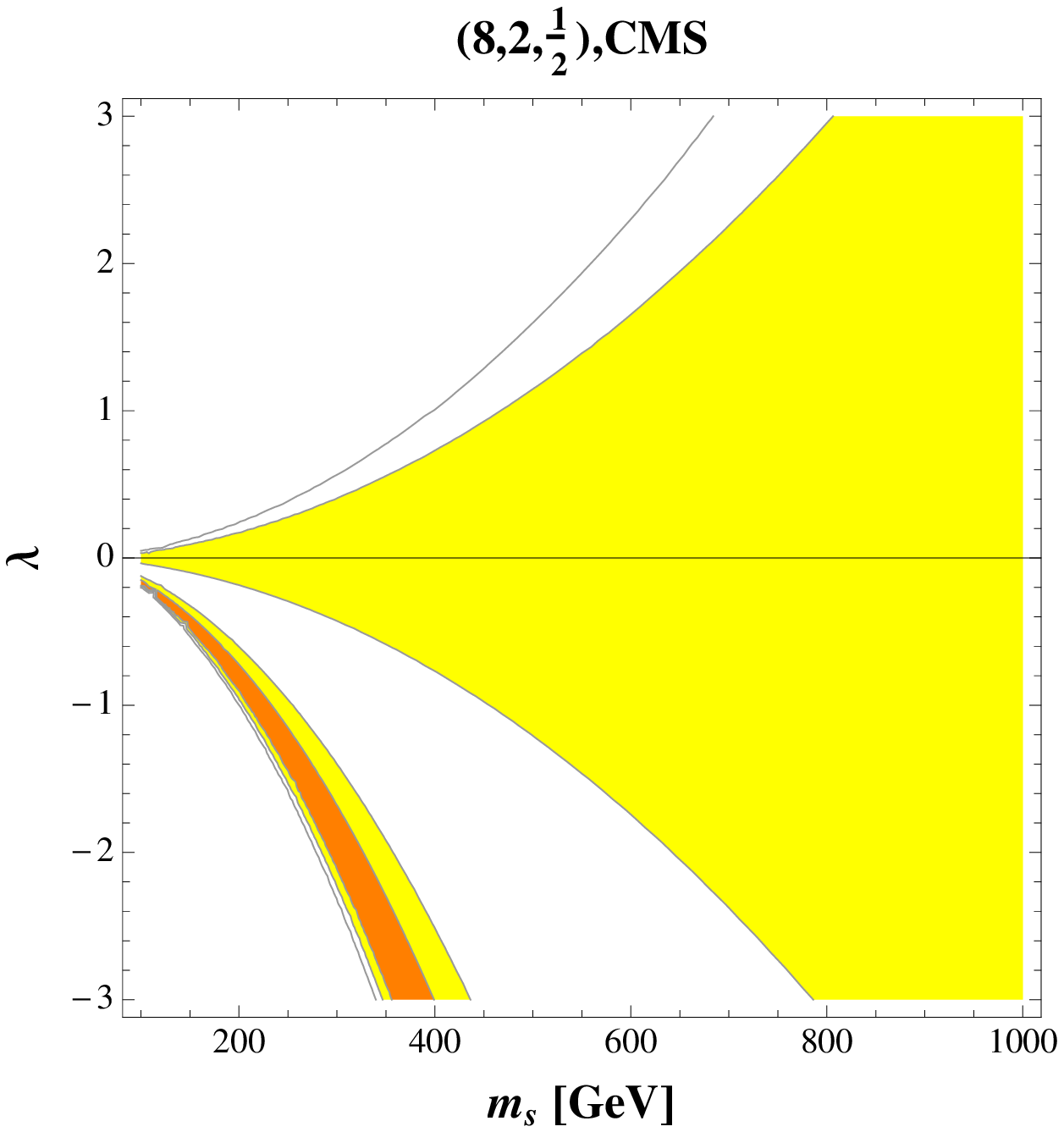}
}
\caption{  
{\bf Scenario A:} Constraints on the coupling $\lambda$ and the mass $m_S$ of the colored scalars. In the left panels, we show constraints coming from ATLAS with the Higgs mass fixed to be 126 GeV, while in the right panels we show  constraints coming from CMS with the Higgs mass fixed to be 124 GeV.
The $1\sigma$ (orange) and the $2\sigma$ (yellow)  allowed regions are presented for the representations $({\bf 3}, {\bf 1}, -\frac{4}{3})$ (top pannels) and $({\bf 8}, {\bf 2}, \frac{1}{2})$ (bottom panels). Finally the solid lines bound the $3\sigma$ allowed regions.
}
\label{fig:colorfit}
\end{figure}

With the assumption that the Higgs mass is in the 124-126 GeV range, as hinted at by the recent ATLAS and CMS results, there are only two free parameters:
the Higgs portal coupling $\lambda$ and the scalar mass $m_S$. 
We first present in Fig.~\ref{fig:colorfit} the results for two colored scalar representations, a triplet $({\bf 3}, {\bf 1}, -\frac{4}{3})$ and 
an octet $({\bf 8},{\bf 2},\frac{1}{2})$. 
In the left panel we show the fit of the present ATLAS data with a Higgs mass fixed to $m_h=126$ GeV; in the right panel the CMS data with $m_h=124$ GeV. The $1\sigma$ ($2\sigma$) bounds are shown in orange (yellow). The solid lines bound the $3\sigma$ allowed regions. 

From the figure we observe that the SM point $\lambda=0$ cannot fit the ATLAS and CMS data at the $1\sigma$ level. The scalar triplet can reach the $1\sigma$ fit of ATLAS and CMS data in two different regions of parameter space which show a sizable enhancement of the di-photon rate: the one with order one Higgs portal couplings and scalar masses in the range $250\,{\rm{GeV}}\lesssim m_S\lesssim 600$ GeV and the other one with very light scalars ($m_S\lesssim 150$ GeV) in which the negative NP contribution to gluon fusion overwhelms the SM contribution and the resulting Higgs production cross section is SM-like. The octet representation instead gives a good fit of the ATLAS and CMS data only for order one Higgs portal couplings and scalar masses in the range $250\,{\rm{GeV}}\lesssim m_S\lesssim 400$ GeV.

Comparing the fits of the ATLAS data to those of the CMS data, we observe that, contrary to the former, the latter has an intermediate mass region that does not give a good fit of the data (see the white region for $\lambda<0$ in Fig.~\ref{fig:colorfit}). These regions, in fact, predict a very suppressed gluon fusion production cross section that is more compatible with the ATLAS data because of the suppressed best fit value for the $WW$ channel (see Tab.~\ref{tab:fit}).

\begin{figure}[h!]
\centerline{
\includegraphics[width=.48\textwidth]{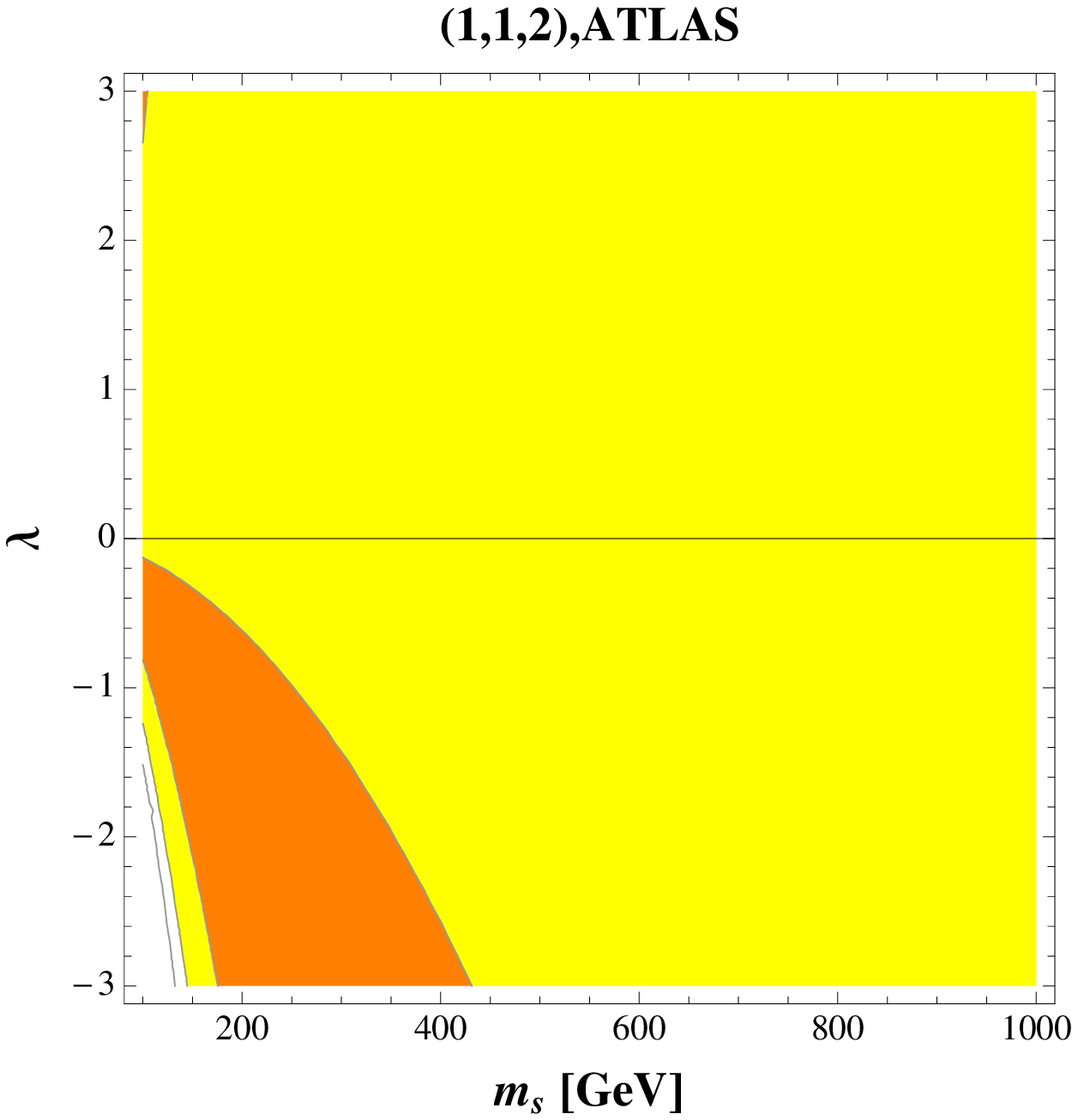}
\includegraphics[width=.48\textwidth]{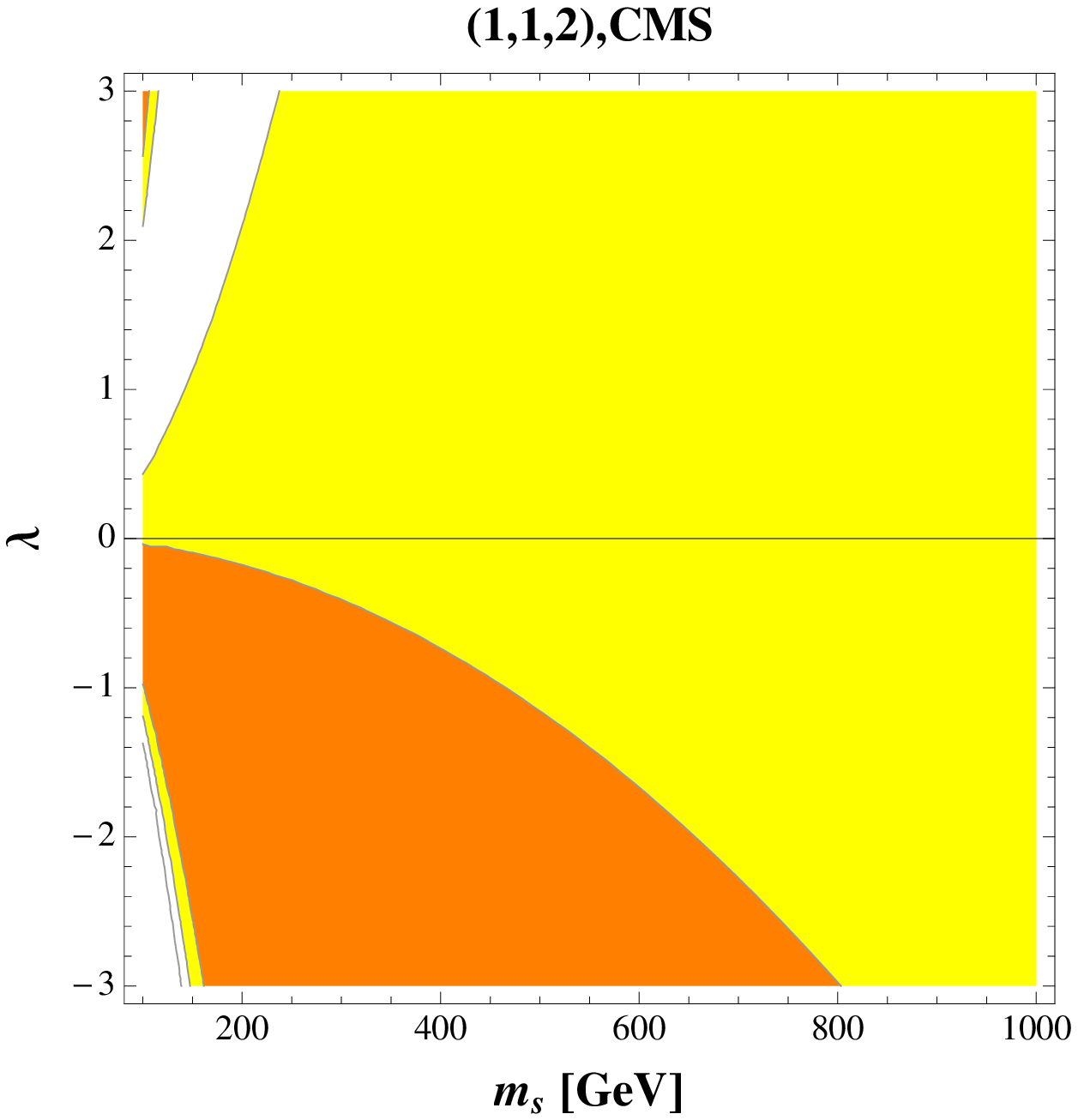}
}
\caption{  
Constraints on the coupling $\lambda$ and the mass $m_S$ of the color neutral scalars. In the left panel, we show constraints coming from ATLAS with the Higgs mass fixed to be 126 GeV, while in the right panel we show  constraints coming from CMS with the Higgs mass fixed to be 124 GeV.
The $1\sigma$ (orange) and the $2\sigma$ (yellow)  allowed regions are presented for the $({\bf 1}, {\bf 1}, 2)$ representation. Finally the solid  line bounds the $3\sigma$ allowed region. }
\label{fig:singlet}
\end{figure}

We next present in   Fig.~\ref{fig:singlet} the fit for the color-singlet scalar $({\bf 1}, {\bf 1}, 2)$. In the left panel we show the fit of the present ATLAS data with a Higgs mass fixed to $m_h=126$ GeV; in the right panel we show the fit of the CMS data with $m_h=124$ GeV. The $1\sigma$ ($2\sigma$) bound is shown in orange (yellow) and the solid line bounds the $3\sigma$ allowed region.  Both fits allow for a sizable region of parameter space at the 1$\sigma$ level ($\lambda<0$).
 
The effects of the color-singlet scalar  are qualitatively different from those of the colored scalars, since only the diphoton rate is modified. The ATLAS and CMS data prefer a slight boost to the $h\rightarrow \gamma \gamma$ rate. As such there are two distinct regions which best fit the data. First, there is a broad region at moderate negative values of $\lambda$ in which the diphoton rate is enhanced. Moving towards positive values of $\lambda$ the rate is reduced as the NP scalar
 begins destructively interfering with the SM $W$ boson loop. However, eventually, for large enough positive values of $\lambda$ and very light scalars, the scalar contribution overwhelms the $W$ contribution to the amplitude and the diphoton rate is enhanced, giving a good fit to the data.  

Besides the illustrative example multiplets presented here, we have studied many other representations in Table~\ref{tab:reps}. Since the results are qualitatively similar to those presented above, we will just make a couple of remarks. In particular, color sextets tend to lead to similar preferred regions as color octets, due to their similar Casimirs ($C({\bf 6}) = 5/2$ vs. $C({\bf 8}) = 3$), and dimensions ($d({\bf 6}) = 6$ vs. $d({\bf 8}) = 8$). Furthermore, multiplets having components with larger electric charges tend to give a good fit to the data for larger scalar masses, while those having components with smaller electric charges require very light scalar masses to fit the data. 

\vspace{5pt}
\noindent {\bf Scenario B (Hidden heavy Higgs)}:
\vspace{5pt}

We now consider the hypothesis that the Higgs boson is heavy $m_h \geq 130$ GeV. 

\begin{figure}[h!]
\centerline{
\includegraphics[width=.46\textwidth]{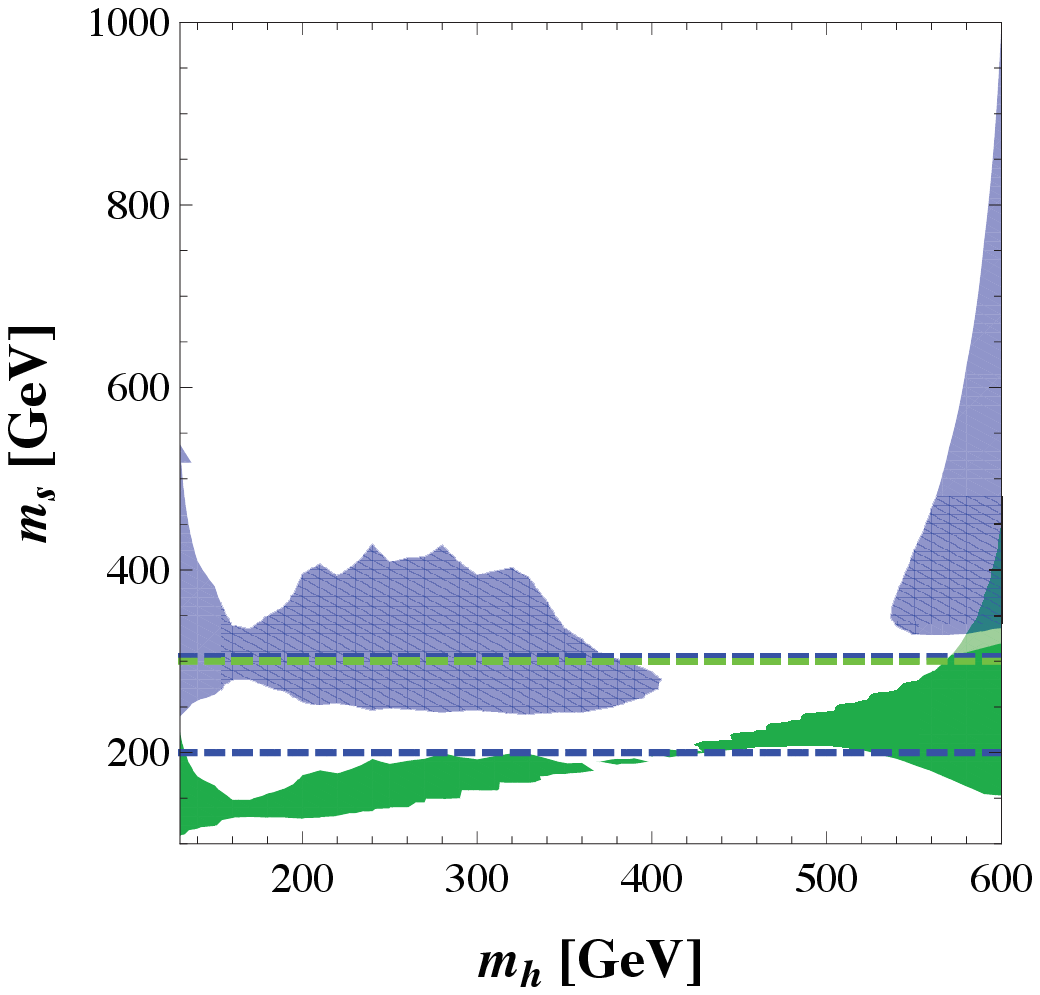}
\includegraphics[width=.45\textwidth]{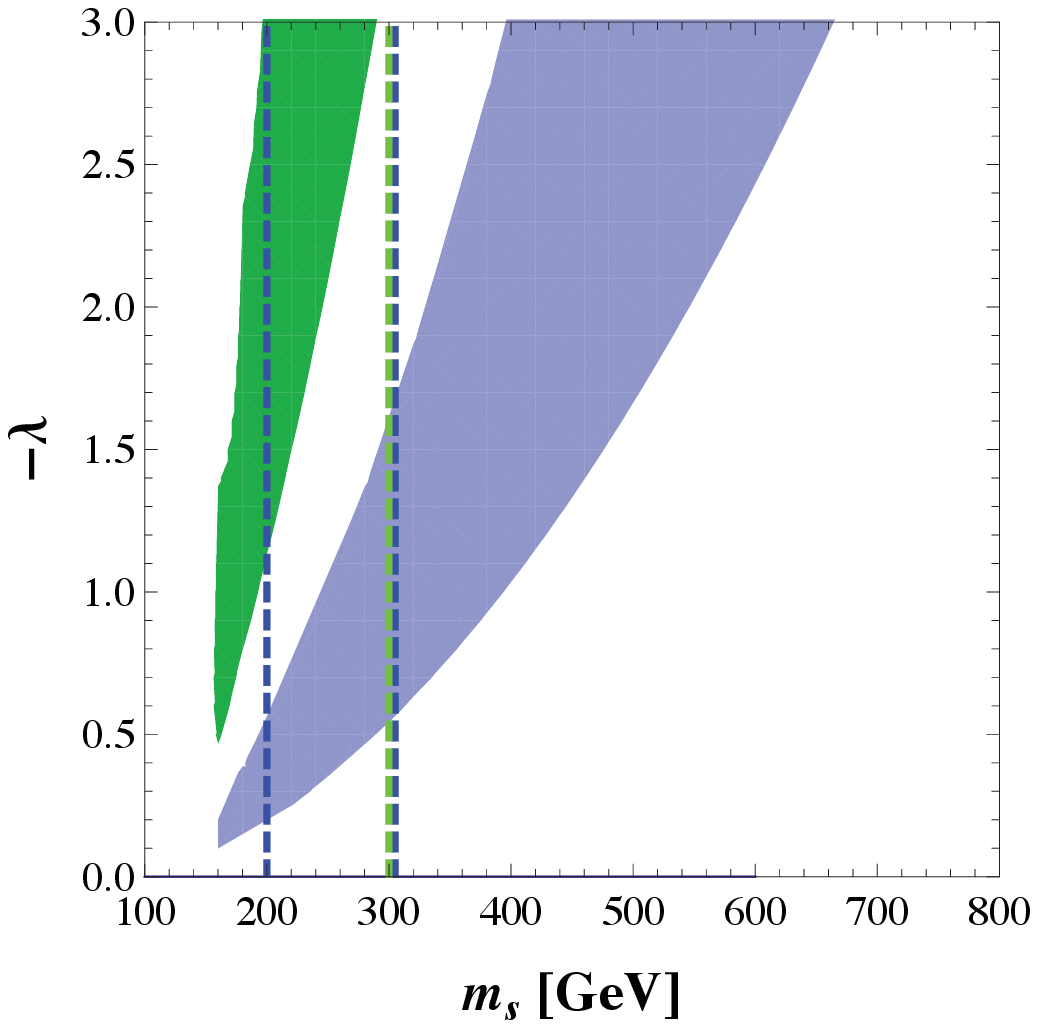}
}
\caption{  
Regions of scalar parameter space that allow a heavy Higgs boson to be consistent with existing LHC searches.
In the left panel we present the allowed region in the $m_h - m_S$ plane with $\lambda = -1$,  while in the right panel we present the region in the $m_S - \lambda$ plane for a 300 GeV Higgs boson. 
We show the results for two representations: $({\bf 3}, {\bf 2}, \frac{1}{6})$ (green) and $({\bf 8}, {\bf 2}, \frac{1}{2})$ (blue). We also show with
dashed lines the corresponding estimate of the current LHC bounds from $(jj)(jj)$ searches: color octets between 200 and 300 GeV and color triplets below 300 GeV are still allowed (see next Section for details).}
\label{fig:triplet_1}
\end{figure}

The left panel in Fig.~\ref{fig:triplet_1} shows the parameter space in the $m_h - m_S$ plane, with Higgs portal coupling fixed to be $\lambda = -1$, in which the Higgs boson would not yet have been observed. 
We consider the cases of a color triplet $({\bf 3}, {\bf 2}, \frac{1}{6})$ (green) and a color octet  $({\bf 8},{\bf 2},\frac{1}{2})$ (blue). For this value of the Higgs portal coupling, we observe that the mass of the triplet representation must be rather light: only a triplet with $120~{\rm{GeV}}\lesssim m_S\lesssim 250$ GeV can reduce the production cross section enough to open up the possibility of having a Higgs in the mass range $130~{\rm{GeV}}\leq m_h\leq 500$ GeV.  Instead, for Higgs between 500 and 600 GeV the scalars must be a bit heavier: $250\,{\rm{GeV}}\lesssim m_S\lesssim 400$ GeV. Due to its larger color factors, the constraint on the mass of the scalar octet is instead weaker, and octets with a mass $250~{\rm{GeV}}\lesssim m_S\lesssim 450$ GeV allow the presence of a Higgs in the entire mass range except in the window $400~{\rm{GeV}}\lesssim m_S\lesssim 520$ GeV in which the constraint from the ATLAS $ZZ$ channel is too strong. We note that we do not have an allowed region for very light scalar masses, in spite of the channel $h\to SS$ being open. In fact, in this region the branching to SM states is still sizeable, and furthermore the negative contribution to the gluon fusion amplitude overwhelms the SM contribution resulting in a production cross section that is much larger than the SM one.

The right panel in Fig.~\ref{fig:triplet_1} shows the parameter space in the $m_S - \lambda$ plane assuming a 300 GeV Higgs boson. 
As one expects, for $\lambda<0$, smaller values of 
$|\lambda|$ require lighter scalar particles to sufficiently reduce the Higgs boson production cross section.  

Finally, we also show in both plots (dashed line)
the current LHC bounds from $(jj)(jj)$ search. This is one of the most generic signatures of scalar pair production in these models (see the discussion in Section~\ref{sec:Collider}). A color octet with a mass between 200 and 300 GeV, as well as a color triplet with mass below 300 GeV, is still allowed by $(jj)(jj)$ searches, while hiding the Higgs to the LHC.

\section{Collider limits on scalars}
\label{sec:Collider}

High energy colliders, in particular the LHC, have excellent potential to probe the new scalars charged under the SM gauge symmetries. We begin with colored scalars, which can be copiously produced  at the LHC through the pair production process:
\begin{equation}
pp\rightarrow S S^{(*)} .
\label{pair}
\end{equation}
The potential signatures can be classified according to the decay products of the scalars. 
The real colored scalar representations  $({\bf 8}, {\bf 1}, 0)$ and  $({\bf 8}, {\bf 3}, 0)$~\cite{Dobrescu:2011px}
do not have renormalizable couplings to SM fermions, but nonetheless can decay at one-loop to a pair of gauge bosons.  The complex colored scalars with nonzero hypercharge decay via the generic coupling
\be
{\cal L} \supset \eta S \bar \psi^{\rm SM}_1 \psi^{\rm SM}_2 + {\rm h.c.},
\label{eq:coupling}
\ee
where $\psi_{i}^{\rm SM}$ denotes a SM fermion.
Assuming that only one of the renormalizable couplings listed in Table~\ref{tab:reps} is dominant,  there are 20 distinct final states for the colored scalars we should consider. Instead of presenting a comprehensive analysis, we focus on several classes of typical signals. 

Both the overall size of the couplings and the relative importance of these channels are quite model dependent. For example, in models with  order one couplings $\eta$, large flavor violating effects have to be expected unless some symmetry principle is imposed, such as Minimal Flavor Violation (MFV).
Of all the scalar representations we consider, only the  $({\bf 1}, {\bf 2}, \frac{1}{2})$ and $({\bf 8}, {\bf 2}, \frac{1}{2})$ multiplets can have MFV couplings to the SM fermions without introducing additional degrees of freedom~\cite{Manohar:2006ga}. Such a MFV model would then predict that the scalars dominantly decay into  third generation fermions. Other scalar models can be made consistent with the MFV hypothesis if the scalars are promoted to transform under the flavor group~\cite{Arnold:2009ay}.
Alternatively,  it is possible that the couplings $\eta$ in Eq.~(\ref{eq:coupling}) are small enough so that flavor constraints are easily satisfied. Such small couplings can easily have UV completions from higher dimensional operators involving SM singlet spurions. 
Most importantly for our considerations,  the smallness of the scalar coupling $\eta$ to SM fermions will not change the Higgs phenomenology, and furthermore will not change the collider signatures of scalar pair production (\ref{pair}) as long as the scalars decay promptly.  Indeed, a typical scalar decay mediated by the Lagrangian in (\ref{eq:coupling}) is prompt for couplings $\eta  \gtrsim 10^{-7}$. We will also consider the case in which the new physics particles are stable on collider time scales. 

It is also possible that,  with different choices of hypercharge, the scalars still decay through higher dimensional operators. An important condition here is that such light scalars do not live long enough to be constrained by dedicated collider searches (see the  detailed discussion below). The collider signals will then depend on the details of the leading irrelevant operator.  We will not  discuss this possibility further in this work. 

In light of these considerations, we will treat one channel at a time, assuming the branching ratio of the scalar in that channel is close to 1. 
With the goal of providing a qualitative survey of the many phenomenological possibilities 
and of pointing out potentially interesting directions for new searches,  
we will estimate the constraints based on current LHC searches with similar final states, and the potential LHC reach with larger datasets. We emphasize that the majority of the NP states considered here have not been directly searched for at the LHC. 
Since LHC searches typically aim at very different underlying NP targets such as SUSY, our estimates here are necessarily rough. 
However, we note that since the production cross section depends very sensitively on the mass of the NP particles,  $\sigma(pp \rightarrow S S^*)  \propto  m_{S}^{-5\  {\rm to} \ -8}$, a misestimation of the sensitivity to production cross section only weakly shifts the the lower bound on the new particle mass.

\noindent\underline{\it $R$-hadrons}

If the couplings governing the decay of scalars  are small enough ($\eta\lesssim 10^{-7}$), then the scalars will be stable on collider time scales. In this case the signature is a long-lived charged  and hadronizing particle, often referred to as a R-hadron, which has been searched for at the LHC \cite{Aad:2011yf,Aad:2011hz, Khachatryan:2011ts,CMS-PAS-EXO-11-022,HSCPs}. The CMS analysis~\cite{HSCPs} sets the most stringent limits: at 95$\%$ C.L. the mass of stable scalar top quarks are bounded to be above 735 GeV. 
We thus estimate the constraint on stable color octet scalars $({\bf 8}, {\bf 1}, 0)$ ($({\bf 8}, {\bf 2}, Y)$) to be approximately at 900 (1000) GeV. Similarly, the constraint on a stable color triplet scalar is about 700 (800) GeV for $({\bf 3}, {\bf 1}, Y)$ ($({\bf 3}, {\bf 2}, Y)$). Therefore, we see that if the scalars are stable on collider time scales, current R-hadron searches already rule out the possibility that a color triplet (octet) plays a role in hiding a heavy Higgs (see Fig.~\ref{fig:triplet_1}).

\noindent\underline{ $(jj)(jj)$}

A common signature for new colored scalars  is the $(jj)(jj)$ final state, which can come from triplet, sextet and octet representations. This is a very difficult signal to uncover due to the large QCD background. For example,  this channel can be used to search for a complex sgluon \cite{Kilic:2008ub,Schumann:2011ji,Bai:2011mr,Choi:2008ub,Kribs:2007ac,Plehn:2008ae}. In a recent ATLAS analysis~\cite{Aad:2011yh} based on 34 pb$^{-1}$, a 
complex scalar gluon is ruled out between 100 and 185 GeV, which translates into an approximate upper bound of 160 GeV on the mass of $({\bf 8}, {\bf 1}, 0)$ (see also \cite{Dobrescu:2011aa}).  
The constraint on the sextet is similar to the one of the octet.
 The production rate of a sextet is approximately double the one of real octet, which roughly translates into a $(10 - 15) \% $ higher limit on the mass of the sextet.   
Finally, the ATLAS search with 34 pb$^{-1}$ puts no constraints on the triplet scalar, but the current luminosity (again assuming the sensitivity to the signal rate goes up by a factor of 10) could be used to constrain the triplet mass scale to be larger than $\sim 150$ GeV. 

Another search for the $(jj)(jj)$ final state based on 2.2 fb$^{-1}$ has been performed by CMS in the higher mass range $300~{\rm{GeV}}<m_S<1200$ GeV~\cite{coloron}, and places bounds on colorons in the range 300-580 GeV. The limits will be similar for the $({\bf 8}, {\bf 1}, 0)$ scalar, while for the $({\bf 8}, {\bf 2}, Y)$ the limit will be close to 900 GeV. For color triplets the limits will be somewhat weaker:  in the 300-500 GeV range for  $({\bf 3}, {\bf 1}, Y)$ and in the 300-550 GeV range for $({\bf 3}, {\bf 2}, Y)$. 

Note that the two existing analyses leaves open the mass window between $\sim 200-300$ GeV that is of central importance for hiding the Higgs with Higgs portal couplings of the order one (see Fig.~\ref{fig:triplet_1}) or to have a good fit of the ATLAS and CMS data for a Higgs at (124-126) GeV (see Fig~\ref{fig:colorfit}). 

\noindent\underline{$(\ell^- j)(\ell^+ j)$}

Another possible final state is $(\ell^- j)(\ell^+ j)$, often referred to as  lepto-quark.  Existing searches at both ATLAS and CMS significantly constrain the lepto-quark parameter space~\cite{Aad:2011uv,Khachatryan:2010mp,Khachatryan:2010mq,Aad:2012cy}.  In particular, an analysis based on 1.03 fb$^{-1}$~\cite{Aad:2012cy} constrains scalar second generation lepto-quarks to be heavier than 685 GeV. Therefore,  for the color triplet to play an interesting role in modifying Higgs phenomenology, its lepto-quark-like couplings should be suppressed. Naive scaling of the sensitivity with luminosity indicates that the mass  reach of the triplet scalar for 10 fb$^{-1}$ would be about 850 GeV. 
 
 \noindent\underline{$jj+\!\not \!\! E_T $}
 
As shown in Tab.~\ref{tab:reps}, the $jj+\!\not \!\!E_T$ signature is possible for color triplet scalars. 
Jets plus missing energy is also a classic signature of the MSSM; as such there exists stringent bounds from LHC searches.  Although not a precise search for the scalars of our interest, as the missing energy in SUSY is carried away by 2 massive LSPs, we adopt this limit as an approximate estimate. ATLAS has preliminary bounds based on $\sim$ 5 fb$^{-1}$ ~\cite{ATLAS-CONF-2012-033} that put a lower bound on the common mass of the first two generation squarks in the 1.2 TeV range (in the limit of a heavy gluino and massles LSP).  Therefore, the constraints on the $({\bf 3}, {\bf 1}, Y)$ scalar, which corresponds to a single squark, is roughly around 1 TeV. 
 
Since the missing energy is due to neutrinos,  $SU(2)_L$ gauge invariance requires that this final state co-exists with a lepto-quark-like final states. Thus, this type of triplet scalar will have constraints and reach similar to those of the lepto-quarks. 
 
\noindent\underline{$t \bar{t} (b \bar{b}) +\!\not \!\! E_T $}
 
Heavy flavor produced in association with missing energy is a possible signal for color triplet scalars. The same final state arises from pair production of top superpartners at the LHC \cite{Meade:2006dw,Matsumoto:2006ws,Kong:2007uu,Han:2008gy,Plehn:2010st,Alwall:2010jc}.
A recent ATLAS search \cite{Aad:2011cw} leads to a lower bound on the stop mass to be around 300 GeV. Hence, in this channel,  the LHC has the potential to fully explore the parameter region of color triplet scalar which is most relevant for Higgs phenomenology. 
 
Due to $SU(2)_L$ gauge invariance, this final state will co-exist with the $t \bar{t}(b \bar{b})+\ell^+ \ell^-$ final state. Multi-lepton  plus multi-jet ($b$-jet) SUSY searches are sensitive to such final states. In particular, this signature is somewhat similar to the $t \bar{t}t \bar{t}$ signal. As  we will discuss in more detail below, the current SUSY search in this channel bounds the  gluino mass to be heavier than about 800 GeV. Taking into account the differences in the production rates for the gluino and the  color triplet scalar, we estimate the bound to be about 500 GeV for the color triplet scalar. 
 
The $b \bar{b} + \!\not \!\! E_T$ final state is also the signal used in the SUSY sbottom searches. As shown in Table~\ref{tab:reps}, whether or not this channel coexists with the $t \bar{t} + \!\not \!\! E_T  $ is model dependent.   The current searches \cite{Aad:2011wc} put an upper bound of about 390 GeV on the sbottom mass. 
At the same time, SU(2)$_L$ gauge invariance requires the presence of the $b \bar{b} + \ell^+ \ell^-$ final state. Therefore, the stronger constraint from lepto-quarks will apply in this case.  

 \noindent\underline{$t \bar{t}t \bar{t}$ and channels with  $t \leftrightarrow b$}
 
 This is a possible class of signals for the color octets and sextets. Searches for similar final states at the LHC  has been recently studied in various NP scenarios~\cite{Lillie:2007hd,Gregoire:2008mr,Chen:2008hh,Acharya:2009gb,Kumar:2009vs,Gregoire:2011ka,Kane:2011zd,Essig:2011qg,Brust:2011tb,Berger:2011af}. 
In particular, this search is similar to the SUSY search channel $\tilde{g} \tilde{g} \to t\bar{t}t\bar{t} + \!\not \!\! E_T$ \cite{ATLAS-CONF-2012-003,CMS-PAS-SUS-11-020}. 
The bound for the gluino from same-sign dileptons \cite{CMS-PAS-SUS-11-020} is around 800 GeV. Therefore, the bound on the scalar octet 
$({\bf 8}, {\bf 2}, Y)$) will be ina similar range . 
 Note that for the representations we consider, the branching ratio to $t \bar{t}t \bar{t}$ is always less than 1.  Since the main strength of this bound comes from the same-sign dilepton signal, the limit will thus be somewhat degraded. A search in the channel $\tilde{g} \tilde{g} \to b\bar{b}b\bar{b} + \!\not \!\! E_T$ \cite{ATLAS-CONF-2012-003} gives a similar constraint. 
We expect with the 7-8 TeV run the LHC will extend this limit to be about 1 TeV, with a similar reach for the octet.

 \noindent\underline{Color neutral scalars}

Finally, we briefly comment on the collider searches for the color neutral scalars. They would typically decay into $W/Z$ as well as SM quark and leptons. Due to small production rate and large SM background from both QCD and EW processes, the search for such NP states is challenging.  For example, some of the signals are similar to SUSY searches for charginos, neutralinos, and sleptons. The limits from  Tevatron and LHC from direct production of such states are weak, typically $<$ 200 GeV, see for example \cite{Bai:2011aa,CDFtrilepton}.

\section{Conclusion}
\label{sec:Conclusion}

The search for the Higgs boson and the determination of its properties are among the central scientific goals of the experimental program at the LHC. Remarkably, being the only elementary scalar field in the SM, the Higgs provides a unique and minimal way of coupling to possible NP states through the operator $H^\dagger H {\mathcal O}_{\rm NP}$, dubbed the Higgs portal. 

Recently at the CERN council, both ATLAS and CMS presented results based on $\sim $ 5 fb$^{-1}$ that significantly advanced the frontier of the the Higgs boson searches. Indeed, these searches rule out a SM-like heavy Higgs in the range of Higgs masses $\sim(130-600)$ GeV,  and are now approaching the sensitivity needed to  detect a light SM-like Higgs. It is therefore quite intriguing that hints of a signal have appeared in both experiments in the small mass window around 124-126 GeV.

In this paper, we have examined the implications of the new data for exotic matter coupling via  the Higgs portal. We have focused on two possible interpretations of the Higgs searches: 1) a Higgs boson in the mass range of 124-126 GeV and 2) a ``hidden" heavy Higgs boson with mass greater than 130 GeV that is underproduced due to the suppression of its gluon fusion production cross section. 

We first performed a model independent analysis on the possible sizes of the Higgs-gluon-gluon and Higgs-photon-photon operators induced by integrating out NP states. The results of this analysis have broad utility and can easily be applied to a variety of NP scenarios that modify Higgs couplings.
In the case of the light Higgs scenario, the existing data already strongly constrains the NP contributions to the Higgs-gluon-gluon coupling,  while there is still sizable room for NP affects on the Higgs-photon-photon coupling. On the other hand, a sizable NP contribution that destructively interferes with the SM $gg\rightarrow h$ amplitude can suppress the gluon fusion production cross section and thus allow the presence of a heavy Higgs boson.

Secondly, we considered concrete UV completions of such higher dimensional operators, in the form of new exotic scalar particles coupling through the Higgs portal. We surveyed a wide range of possible $SU(3)_C \times SU(2)_L \times U(1)_Y$ scalar representations, determining the allowed parameter space from the current data. Assuming a 124-126 GeV Higgs boson, for order one Higgs portal couplings, the current data allow the color triplet (octet) to be as light as $\sim 150$ ($\sim 250$) GeV, while color neutral scalars are largely unconstrained. For a heavy Higgs to be hidden,  the color triplet (octet) must have a mass of the order $\sim 200$ GeV ($\sim 300$ GeV) for order one negative Higgs portal couplings.

Such NP particles can also be searched for directly at the LHC. We have investigated the constraints and prospects under the assumption that the new scalars can decay via renormalizable interactions. 
Although some of the parameter space has already been excluded, large regions remain unexplored. In particular, the $4j$ signature is only weakly constrained at low scalar masses due to the large QCD background. Other signatures such as lepto-quark-like or top-rich final states, if available, lead to more stringent bounds. Moreover, with dedicated searches a significant portion of the most interesting parameter regions for Higgs phenomenolgy can be probed with the current and future datasets. Indeed, if deviations from a SM-like Higgs are observed, new light exotic matter coupling through the Higgs portal will provide a promising and experimentally testable explanation, as such states can be directly observed in the near future at the LHC.

\section{Acknowledgements}
We thank Wolfgang Altmannshofer, David McKeen, Josef Pradler for helpful discussions. We are also grateful to Ian Low for pointing out an incorrect numerical factor in the first version of this paper. Work at ANL is supported in part by the U.S. Department of Energy (DOE), Div. of HEP, Contract DE-AC02-06CH11357. L.T.W. and B.B. are supported by the NSF under grant PHY-0756966 and the DOE Early Career Award under grant DE-SC0003930.

\newpage

\appendix

\section{Loop functions}

In this Appendix we give the definitions of the loop functions used in the text. 
For the scalar, fermion, and vector particles running in the loop, respectively, we have
\begin{eqnarray}
A_{0} (\tau) & = & \tau^{-2} [f(\tau) -\tau   ],    \label{eq:A0} \\
A_{1/2} (\tau) &  = & 2 \tau^{-2} [\tau +(\tau -1) f(\tau) ],  \label{eq:A12} \\
A_1(\tau) & = & 2 + \frac{3}{\tau} + \frac{3}{\tau} (2 - \frac{1}{\tau}) f(\tau)   \label{eq:A1} 
\end{eqnarray}
where $\tau = m_h^2/(4 m^2)$, with $m$ the mass of the particle in the loop,
and the function $f(\tau)$ defined as
\begin{equation}
\label{ftau}
f(\tau) = 
\begin{cases}
\displaystyle{\arcsin^2 \sqrt{\tau} }\qquad \qquad\qquad\qquad \qquad ~\,\, \,\,  {\rm for }\quad \tau \leq 1, \\
\displaystyle{-\frac{1}{4} \left[\log \frac{1+ \sqrt{1-\tau^{-1} } }{   1 - \sqrt{1-\tau^{-1} } } - i \pi \right]^2} 
 \qquad ~~{\rm for }\quad \tau > 1.
\end{cases}
\end{equation}

\bibliographystyle{jhep}
\bibliography{papbib}

\providecommand{\href}[2]{#2}\begingroup\raggedright\begin{thebibliography}{100}

\bibitem{Higgs:1964ia}
P.~W. Higgs, {\it {Broken symmetries, massless particles and gauge fields}},
  {\em Phys.Lett.} {\bf 12} (1964) 132--133.

\bibitem{Englert:1964et}
F.~Englert and R.~Brout, {\it {Broken Symmetry and the Mass of Gauge Vector
  Mesons}},  {\em Phys.Rev.Lett.} {\bf 13} (1964) 321--322.

\bibitem{Higgs:1964pj}
P.~W. Higgs, {\it {Broken Symmetries and the Masses of Gauge Bosons}},  {\em
  Phys.Rev.Lett.} {\bf 13} (1964) 508--509.

\bibitem{Guralnik:1964eu}
G.~Guralnik, C.~Hagen, and T.~Kibble, {\it {Global Conservation Laws and
  Massless Particles}},  {\em Phys.Rev.Lett.} {\bf 13} (1964) 585--587.

\bibitem{Weinberg:1967tq}
S.~Weinberg, {\it {A Model of Leptons}},  {\em Phys.Rev.Lett.} {\bf 19} (1967)
  1264--1266.

\bibitem{Djouadi:2005gi}
A.~Djouadi, {\it {The Anatomy of electro-weak symmetry breaking. I: The Higgs
  boson in the standard model}},  {\em Phys.Rept.} {\bf 457} (2008) 1--216,
  [\href{http://xxx.lanl.gov/abs/hep-ph/0503172}{{\tt hep-ph/0503172}}].

\bibitem{ATLASHiggs}
{\bf ATLAS} Collaboration, ``An update to the combined search for the standard
  model higgs boson with the atlas detector at the lhc using up to 4.9
  fb$^{−1}$ of pp collision data at sqrt(s) = 7 tev.''
  \url{http://cdsweb.cern.ch/record/1430033}.

\bibitem{CMSHiggs}
{\bf CMS} Collaboration, S.~Chatrchyan {\em et.~al.}, {\it {Combined results of
  searches for the standard model Higgs boson in pp collisions at sqrt(s) = 7
  TeV}},  \href{http://xxx.lanl.gov/abs/1202.1488}{{\tt arXiv:1202.1488}}.

\bibitem{Ibe:2011aa}
M.~Ibe and T.~T. Yanagida, {\it {The Lightest Higgs Boson Mass in Pure Gravity
  Mediation Model}},  \href{http://xxx.lanl.gov/abs/1112.2462}{{\tt
  arXiv:1112.2462}}.

\bibitem{Arbey:2011ab}
A.~Arbey, M.~Battaglia, A.~Djouadi, F.~Mahmoudi, and J.~Quevillon, {\it
  {Implications of a 125 GeV Higgs for supersymmetric models}},
  \href{http://xxx.lanl.gov/abs/1112.3028}{{\tt arXiv:1112.3028}}.

\bibitem{Heinemeyer:2011aa}
S.~Heinemeyer, O.~Stal, and G.~Weiglein, {\it {Interpreting the LHC Higgs
  Search Results in the MSSM}},  \href{http://xxx.lanl.gov/abs/1112.3026}{{\tt
  arXiv:1112.3026}}.

\bibitem{Li:2011ab}
T.~Li, J.~A. Maxin, D.~V. Nanopoulos, and J.~W. Walker, {\it {A Higgs Mass
  Shift to 125 GeV and A Multi-Jet Supersymmetry Signal: Miracle of the
  Flippons at the sqrt{s} = 7 TeV LHC}},
  \href{http://xxx.lanl.gov/abs/1112.3024}{{\tt arXiv:1112.3024}}.

\bibitem{EliasMiro:2011aa}
J.~Elias-Miro, J.~R. Espinosa, G.~F. Giudice, G.~Isidori, A.~Riotto, {\em
  et.~al.}, {\it {Higgs mass implications on the stability of the electroweak
  vacuum}},  \href{http://xxx.lanl.gov/abs/1112.3022}{{\tt arXiv:1112.3022}}.

\bibitem{Feng:2011aa}
J.~L. Feng, K.~T. Matchev, and D.~Sanford, {\it {Focus Point Supersymmetry
  Redux}},  \href{http://xxx.lanl.gov/abs/1112.3021}{{\tt arXiv:1112.3021}}.

\bibitem{Baer:2011ab}
H.~Baer, V.~Barger, and A.~Mustafayev, {\it {Implications of a 125 GeV Higgs
  scalar for LHC SUSY and neutralino dark matter searches}},
  \href{http://xxx.lanl.gov/abs/1112.3017}{{\tt arXiv:1112.3017}}.

\bibitem{Englert:2011aa}
C.~Englert, T.~Plehn, M.~Rauch, D.~Zerwas, and P.~M. Zerwas, {\it {LHC:
  Standard Higgs and Hidden Higgs}},
  \href{http://xxx.lanl.gov/abs/1112.3007}{{\tt arXiv:1112.3007}}.

\bibitem{Carena:2011aa}
M.~Carena, S.~Gori, N.~R. Shah, and C.~E. Wagner, {\it {A 125 GeV SM-like Higgs
  in the MSSM and the $\gamma \gamma$ rate}},
  \href{http://xxx.lanl.gov/abs/1112.3336}{{\tt arXiv:1112.3336}}.

\bibitem{Djouadi:2011aa}
A.~Djouadi, O.~Lebedev, Y.~Mambrini, and J.~Quevillon, {\it {Implications of
  LHC searches for Higgs--portal dark matter}},
  \href{http://xxx.lanl.gov/abs/1112.3299}{{\tt arXiv:1112.3299}}.

\bibitem{Ferreira:2011aa}
P.~Ferreira, R.~Santos, M.~Sher, and J.~P. Silva, {\it {Implications of the LHC
  two-photon signal for two-Higgs-doublet models}},
  \href{http://xxx.lanl.gov/abs/1112.3277}{{\tt arXiv:1112.3277}}.

\bibitem{Guo:2011ab}
G.~Guo, B.~Ren, and X.-G. He, {\it {LHC Evidence Of A 126 GeV Higgs Boson From
  $H \to \gamma \gamma$ With Three And Four Generations}},
  \href{http://xxx.lanl.gov/abs/1112.3188}{{\tt arXiv:1112.3188}}.

\bibitem{Moroi:2011aa}
T.~Moroi, R.~Sato, and T.~T. Yanagida, {\it {Extra Matters Decree the
  Relatively Heavy Higgs of Mass about 125 GeV in the Supersymmetric Model}},
  \href{http://xxx.lanl.gov/abs/1112.3142}{{\tt arXiv:1112.3142}}.

\bibitem{Moroi:2011ab}
T.~Moroi and K.~Nakayama, {\it {Wino LSP detection in the light of recent Higgs
  searches at the LHC}},  \href{http://xxx.lanl.gov/abs/1112.3123}{{\tt
  arXiv:1112.3123}}.

\bibitem{Xing:2011aa}
Z.-z. Xing, H.~Zhang, and S.~Zhou, {\it {Impacts of the Higgs mass on vacuum
  stability, running fermion masses and two-body Higgs decays}},
  \href{http://xxx.lanl.gov/abs/1112.3112}{{\tt arXiv:1112.3112}}.

\bibitem{Draper:2011aa}
P.~Draper, P.~Meade, M.~Reece, and D.~Shih, {\it {Implications of a 125 GeV
  Higgs for the MSSM and Low-Scale SUSY Breaking}},
  \href{http://xxx.lanl.gov/abs/1112.3068}{{\tt arXiv:1112.3068}}.

\bibitem{Cheung:2011aa}
C.~Cheung and Y.~Nomura, {\it {Higgs Descendants}},
  \href{http://xxx.lanl.gov/abs/1112.3043}{{\tt arXiv:1112.3043}}.

\bibitem{Hall:2011aa}
L.~J. Hall, D.~Pinner, and J.~T. Ruderman, {\it {A Natural SUSY Higgs Near 126
  GeV}},  \href{http://xxx.lanl.gov/abs/1112.2703}{{\tt arXiv:1112.2703}}.

\bibitem{Kane:2011kj}
G.~Kane, P.~Kumar, R.~Lu, and B.~Zheng, {\it {Higgs Mass Prediction for
  Realistic String/M Theory Vacua}},
  \href{http://xxx.lanl.gov/abs/1112.1059}{{\tt arXiv:1112.1059}}.

\bibitem{Kadastik:2011aa}
M.~Kadastik, K.~Kannike, A.~Racioppi, and M.~Raidal, {\it {Implications of 125
  GeV Higgs boson on scalar dark matter and on the CMSSM phenomenology}},
  \href{http://xxx.lanl.gov/abs/1112.3647}{{\tt arXiv:1112.3647}}.

\bibitem{Akula:2011aa}
S.~Akula, B.~Altunkaynak, D.~Feldman, P.~Nath, and G.~Peim, {\it {Higgs Boson
  Mass Predictions in SUGRA Unification and Recent LHC-7 Results}},
  \href{http://xxx.lanl.gov/abs/1112.3645}{{\tt arXiv:1112.3645}}.

\bibitem{Buchmueller:2011ab}
O.~Buchmueller, R.~Cavanaugh, A.~De~Roeck, M.~Dolan, J.~Ellis, {\em et.~al.},
  {\it {Higgs and Supersymmetry}},
  \href{http://xxx.lanl.gov/abs/1112.3564}{{\tt arXiv:1112.3564}}.

\bibitem{Masina:2011aa}
I.~Masina and A.~Notari, {\it {The Higgs mass range from Standard Model false
  vacuum Inflation in scalar-tensor gravity}},
  \href{http://xxx.lanl.gov/abs/1112.2659}{{\tt arXiv:1112.2659}}.

\bibitem{Masina:2011un}
I.~Masina and A.~Notari, {\it {Standard Model false vacuum Inflation:
  correlating the tensor-to-scalar ratio to the top and Higgs masses}},
  \href{http://xxx.lanl.gov/abs/1112.5430}{{\tt arXiv:1112.5430}}.

\bibitem{Ellwanger:2011aa}
U.~Ellwanger, {\it {A Higgs boson near 125 GeV with enhanced di-photon signal
  in the NMSSM}},  \href{http://xxx.lanl.gov/abs/1112.3548}{{\tt
  arXiv:1112.3548}}.

\bibitem{Cheung:2011nv}
K.~Cheung and T.-C. Yuan, {\it {Could the excess seen at 124-126 GeV be due to
  the Randall-Sundrum Radion?}},  {\em Phys.Rev.Lett.} {\bf 108} (2012) 141602,
  [\href{http://xxx.lanl.gov/abs/1112.4146}{{\tt arXiv:1112.4146}}].

\bibitem{deSandes:2011zs}
H.~de~Sandes and R.~Rosenfeld, {\it {Radion-Higgs mixing effects on bounds from
  LHC Higgs Searches}},  {\em Phys.Rev.} {\bf D85} (2012) 053003,
  [\href{http://xxx.lanl.gov/abs/1111.2006}{{\tt arXiv:1111.2006}}].

\bibitem{Baek:2011aa}
S.-W. Baek, P.~Ko, and W.-I. Park, {\it {Search for the Higgs portal to a
  singlet fermionic dark matter at the LHC}},  {\em JHEP} {\bf 1202} (2012)
  047, [\href{http://xxx.lanl.gov/abs/1112.1847}{{\tt arXiv:1112.1847}}].

\bibitem{Arbey:2011aa}
A.~Arbey, M.~Battaglia, and F.~Mahmoudi {\em Eur.Phys.J.} {\bf C72} (2012)
  1906, [\href{http://xxx.lanl.gov/abs/1112.3032}{{\tt arXiv:1112.3032}}].

\bibitem{Silveira:1985rk}
V.~Silveira and A.~Zee, {\it {SCALAR PHANTOMS}},  {\em Phys.Lett.} {\bf B161}
  (1985) 136.

\bibitem{McDonald:1993ex}
J.~McDonald, {\it {Gauge singlet scalars as cold dark matter}},  {\em
  Phys.Rev.} {\bf D50} (1994) 3637--3649,
  [\href{http://xxx.lanl.gov/abs/hep-ph/0702143}{{\tt hep-ph/0702143}}].

\bibitem{Burgess:2000yq}
C.~Burgess, M.~Pospelov, and T.~ter Veldhuis, {\it {The Minimal model of
  nonbaryonic dark matter: A Singlet scalar}},  {\em Nucl.Phys.} {\bf B619}
  (2001) 709--728, [\href{http://xxx.lanl.gov/abs/hep-ph/0011335}{{\tt
  hep-ph/0011335}}].

\bibitem{Berger:2002vs}
E.~L. Berger, C.-W. Chiang, J.~Jiang, T.~M. Tait, and C.~E. Wagner, {\it {Higgs
  boson decay into hadronic jets}},  {\em Phys.Rev.} {\bf D66} (2002) 095001,
  [\href{http://xxx.lanl.gov/abs/hep-ph/0205342}{{\tt hep-ph/0205342}}].

\bibitem{Dermisek:2005ar}
R.~Dermisek and J.~F. Gunion, {\it {Escaping the large fine tuning and little
  hierarchy problems in the next to minimal supersymmetric model and h aa
  decays}},  {\em Phys.Rev.Lett.} {\bf 95} (2005) 041801,
  [\href{http://xxx.lanl.gov/abs/hep-ph/0502105}{{\tt hep-ph/0502105}}].

\bibitem{Chang:2005ht}
S.~Chang, P.~J. Fox, and N.~Weiner, {\it {Naturalness and Higgs decays in the
  MSSM with a singlet}},  {\em JHEP} {\bf 0608} (2006) 068,
  [\href{http://xxx.lanl.gov/abs/hep-ph/0511250}{{\tt hep-ph/0511250}}].

\bibitem{Bellazzini:2009xt}
B.~Bellazzini, C.~Csaki, A.~Falkowski, and A.~Weiler, {\it {Buried Higgs}},
  {\em Phys.Rev.} {\bf D80} (2009) 075008,
  [\href{http://xxx.lanl.gov/abs/0906.3026}{{\tt arXiv:0906.3026}}].

\bibitem{Falkowski:2010cm}
A.~Falkowski, J.~T. Ruderman, T.~Volansky, and J.~Zupan, {\it {Hidden Higgs
  Decaying to Lepton Jets}},  {\em JHEP} {\bf 1005} (2010) 077,
  [\href{http://xxx.lanl.gov/abs/1002.2952}{{\tt arXiv:1002.2952}}].

\bibitem{Kanemura:2010sh}
S.~Kanemura, S.~Matsumoto, T.~Nabeshima, and N.~Okada, {\it {Can WIMP Dark
  Matter overcome the Nightmare Scenario?}},  {\em Phys.Rev.} {\bf D82} (2010)
  055026, [\href{http://xxx.lanl.gov/abs/1005.5651}{{\tt arXiv:1005.5651}}].

\bibitem{Englert:2011us}
C.~Englert, J.~Jaeckel, E.~Re, and M.~Spannowsky, {\it {Evasive Higgs Maneuvers
  at the LHC}},  {\em Phys.Rev.} {\bf D85} (2012) 035008,
  [\href{http://xxx.lanl.gov/abs/1111.1719}{{\tt arXiv:1111.1719}}]. 10 pages,
  6 figures, references added, final version accepted for publication in PRD.

\bibitem{Chang:2008cw}
S.~Chang, R.~Dermisek, J.~F. Gunion, and N.~Weiner, {\it {Nonstandard Higgs
  Boson Decays}},  {\em Ann.Rev.Nucl.Part.Sci.} {\bf 58} (2008) 75--98,
  [\href{http://xxx.lanl.gov/abs/0801.4554}{{\tt arXiv:0801.4554}}].

\bibitem{Falkowski:2010hi}
A.~Falkowski, D.~Krohn, L.-T. Wang, J.~Shelton, and A.~Thalapillil, {\it
  {Unburied Higgs boson: Jet substructure techniques for searching for Higgs'
  decay into gluons}},  {\em Phys.Rev.} {\bf D84} (2011) 074022,
  [\href{http://xxx.lanl.gov/abs/1006.1650}{{\tt arXiv:1006.1650}}].

\bibitem{Schabinger:2005ei}
R.~Schabinger and J.~D. Wells, {\it {A Minimal spontaneously broken hidden
  sector and its impact on Higgs boson physics at the large hadron collider}},
  {\em Phys.Rev.} {\bf D72} (2005) 093007,
  [\href{http://xxx.lanl.gov/abs/hep-ph/0509209}{{\tt hep-ph/0509209}}].

\bibitem{O'Connell:2006wi}
D.~O'Connell, M.~J. Ramsey-Musolf, and M.~B. Wise, {\it {Minimal Extension of
  the Standard Model Scalar Sector}},  {\em Phys.Rev.} {\bf D75} (2007) 037701,
  [\href{http://xxx.lanl.gov/abs/hep-ph/0611014}{{\tt hep-ph/0611014}}].

\bibitem{Barger:2007im}
V.~Barger, P.~Langacker, M.~McCaskey, M.~J. Ramsey-Musolf, and G.~Shaughnessy,
  {\it {LHC Phenomenology of an Extended Standard Model with a Real Scalar
  Singlet}},  {\em Phys.Rev.} {\bf D77} (2008) 035005,
  [\href{http://xxx.lanl.gov/abs/0706.4311}{{\tt arXiv:0706.4311}}].

\bibitem{Carena:2011fc}
M.~Carena, P.~Draper, T.~Liu, and C.~Wagner, {\it {The 7 TeV LHC Reach for MSSM
  Higgs Bosons}},  {\em Phys.Rev.} {\bf D84} (2011) 095010,
  [\href{http://xxx.lanl.gov/abs/1107.4354}{{\tt arXiv:1107.4354}}].

\bibitem{Englert:2011yb}
C.~Englert, T.~Plehn, D.~Zerwas, and P.~M. Zerwas, {\it {Exploring the Higgs
  portal}},  {\em Phys.Lett.} {\bf B703} (2011) 298--305,
  [\href{http://xxx.lanl.gov/abs/1106.3097}{{\tt arXiv:1106.3097}}].

\bibitem{Lebedev:2011iq}
O.~Lebedev, H.~M. Lee, and Y.~Mambrini, {\it {Vector Higgs-portal dark matter
  and the invisible Higgs}},  {\em Phys.Lett.} {\bf B707} (2012) 570--576,
  [\href{http://xxx.lanl.gov/abs/1111.4482}{{\tt arXiv:1111.4482}}]. 8 pages, 8
  figures, Version to appear in Phys. Lett. B.

\bibitem{Draper:2009fh}
P.~Draper, T.~Liu, and C.~E. Wagner, {\it {Prospects for MSSM Higgs Searches at
  the Tevatron}},  {\em Phys.Rev.} {\bf D80} (2009) 035025,
  [\href{http://xxx.lanl.gov/abs/0905.4721}{{\tt arXiv:0905.4721}}].

\bibitem{Draper:2009au}
P.~Draper, T.~Liu, and C.~E. Wagner, {\it {Prospects for Higgs Searches at the
  Tevatron and LHC in the MSSM with Explicit CP-violation}},  {\em Phys.Rev.}
  {\bf D81} (2010) 015014, [\href{http://xxx.lanl.gov/abs/0911.0034}{{\tt
  arXiv:0911.0034}}].

\bibitem{Weihs:2011wp}
E.~Weihs and J.~Zurita, {\it {Dark Higgs Models at the 7 TeV LHC}},  {\em JHEP}
  {\bf 1202} (2012) 041, [\href{http://xxx.lanl.gov/abs/1110.5909}{{\tt
  arXiv:1110.5909}}].

\bibitem{Manohar:2006gz}
A.~V. Manohar and M.~B. Wise, {\it {Modifications to the properties of the
  Higgs boson}},  {\em Phys.Lett.} {\bf B636} (2006) 107--113,
  [\href{http://xxx.lanl.gov/abs/hep-ph/0601212}{{\tt hep-ph/0601212}}].

\bibitem{Arnesen:2008fb}
C.~Arnesen, I.~Z. Rothstein, and J.~Zupan, {\it {Smoking Guns for On-Shell New
  Physics at the LHC}},  {\em Phys.Rev.Lett.} {\bf 103} (2009) 151801,
  [\href{http://xxx.lanl.gov/abs/0809.1429}{{\tt arXiv:0809.1429}}].

\bibitem{Peskin:2001rw}
M.~E. Peskin and J.~D. Wells, {\it {How can a heavy Higgs boson be consistent
  with the precision electroweak measurements?}},  {\em Phys.Rev.} {\bf D64}
  (2001) 093003, [\href{http://xxx.lanl.gov/abs/hep-ph/0101342}{{\tt
  hep-ph/0101342}}].

\bibitem{Djouadi:1998az}
A.~Djouadi, {\it {Squark effects on Higgs boson production and decay at the
  LHC}},  {\em Phys.Lett.} {\bf B435} (1998) 101--108,
  [\href{http://xxx.lanl.gov/abs/hep-ph/9806315}{{\tt hep-ph/9806315}}].

\bibitem{Carena:2002qg}
M.~S. Carena, S.~Heinemeyer, C.~Wagner, and G.~Weiglein, {\it {Suggestions for
  benchmark scenarios for MSSM Higgs boson searches at hadron colliders}},
  {\em Eur.Phys.J.} {\bf C26} (2003) 601--607,
  [\href{http://xxx.lanl.gov/abs/hep-ph/0202167}{{\tt hep-ph/0202167}}].

\bibitem{Dermisek:2007fi}
R.~Dermisek and I.~Low, {\it {Probing the Stop Sector and the Sanity of the
  MSSM with the Higgs Boson at the LHC}},  {\em Phys.Rev.} {\bf D77} (2008)
  035012, [\href{http://xxx.lanl.gov/abs/hep-ph/0701235}{{\tt
  hep-ph/0701235}}].

\bibitem{Falkowski:2007hz}
A.~Falkowski, {\it {Pseudo-goldstone Higgs production via gluon fusion}},  {\em
  Phys.Rev.} {\bf D77} (2008) 055018,
  [\href{http://xxx.lanl.gov/abs/0711.0828}{{\tt arXiv:0711.0828}}].

\bibitem{Low:2010mr}
I.~Low and A.~Vichi, {\it {On the production of a composite Higgs boson}},
  {\em Phys.Rev.} {\bf D84} (2011) 045019,
  [\href{http://xxx.lanl.gov/abs/1010.2753}{{\tt arXiv:1010.2753}}].

\bibitem{Bai:2011aa}
Y.~Bai, J.~Fan, and J.~L. Hewett, {\it {Hiding a Heavy Higgs Boson at the 7 TeV
  LHC}},  \href{http://xxx.lanl.gov/abs/1112.1964}{{\tt arXiv:1112.1964}}.

\bibitem{Dobrescu:2011aa}
B.~A. Dobrescu, G.~D. Kribs, and A.~Martin, {\it {Higgs Underproduction at the
  LHC}},  \href{http://xxx.lanl.gov/abs/1112.2208}{{\tt arXiv:1112.2208}}.

\bibitem{Smith:1982qu}
P.~F. Smith {\em et.~al.}, {\it {A SEARCH FOR ANOMALOUS HYDROGEN IN ENRICHED
  D-2 O, USING A TIME-OF-FLIGHT SPECTROMETER}},  {\em Nucl. Phys.} {\bf B206}
  (1982) 333--348.

\bibitem{HSCPs}
{\bf CMSn} Collaboration, {\it Search for hscps},  Tech. Rep. EXO-11-022, 2012.

\bibitem{DelNobile:2009st}
E.~Del~Nobile, R.~Franceschini, D.~Pappadopulo, and A.~Strumia, {\it {Minimal
  Matter at the Large Hadron Collider}},  {\em Nucl.Phys.} {\bf B826} (2010)
  217--234, [\href{http://xxx.lanl.gov/abs/0908.1567}{{\tt arXiv:0908.1567}}].

\bibitem{HiggsXsection}
\url{https://twiki.cern.ch/twiki/bin/view/LHCPhysics/CrossSections}.

\bibitem{Bonciani:2007ex}
R.~Bonciani, G.~Degrassi, and A.~Vicini, {\it {Scalar particle contribution to
  Higgs production via gluon fusion at NLO}},  {\em JHEP} {\bf 0711} (2007)
  095, [\href{http://xxx.lanl.gov/abs/0709.4227}{{\tt arXiv:0709.4227}}].

\bibitem{Boughezal:2010ry}
R.~Boughezal and F.~Petriello, {\it {Color-octet scalar effects on Higgs boson
  production in gluon fusion}},  {\em Phys.Rev.} {\bf D81} (2010) 114033,
  [\href{http://xxx.lanl.gov/abs/1003.2046}{{\tt arXiv:1003.2046}}].

\bibitem{Dobrescu:2011px}
B.~A. Dobrescu and G.~Z. Krnjaic, {\it {Weak-triplet, color-octet scalars and
  the CDF dijet excess}},  \href{http://xxx.lanl.gov/abs/1104.2893}{{\tt
  arXiv:1104.2893}}.

\bibitem{Manohar:2006ga}
A.~V. Manohar and M.~B. Wise, {\it {Flavor changing neutral currents, an
  extended scalar sector, and the Higgs production rate at the CERN LHC}},
  {\em Phys.Rev.} {\bf D74} (2006) 035009,
  [\href{http://xxx.lanl.gov/abs/hep-ph/0606172}{{\tt hep-ph/0606172}}].

\bibitem{Arnold:2009ay}
J.~M. Arnold, M.~Pospelov, M.~Trott, and M.~B. Wise, {\it {Scalar
  Representations and Minimal Flavor Violation}},  {\em JHEP} {\bf 1001} (2010)
  073, [\href{http://xxx.lanl.gov/abs/0911.2225}{{\tt arXiv:0911.2225}}].

\bibitem{Aad:2011yf}
{\bf ATLAS} Collaboration, G.~Aad {\em et.~al.}, {\it {Search for stable
  hadronising squarks and gluinos with the ATLAS experiment at the LHC}},  {\em
  Phys.Lett.} {\bf B701} (2011) 1--19,
  [\href{http://xxx.lanl.gov/abs/1103.1984}{{\tt arXiv:1103.1984}}]. *
  Temporary entry *.

\bibitem{Aad:2011hz}
{\bf ATLAS} Collaboration, G.~Aad {\em et.~al.}, {\it {Search for Heavy
  Long-Lived Charged Particles with the ATLAS detector in pp collisions at
  sqrt(s) = 7 TeV}},  {\em Phys.Lett.} {\bf B703} (2011) 428--446,
  [\href{http://xxx.lanl.gov/abs/1106.4495}{{\tt arXiv:1106.4495}}]. *
  Temporary entry *.

\bibitem{Khachatryan:2011ts}
{\bf CMS} Collaboration, V.~Khachatryan {\em et.~al.}, {\it {Search for Heavy
  Stable Charged Particles in pp collisions at sqrt(s)=7 TeV}},  {\em JHEP}
  {\bf 1103} (2011) 024, [\href{http://xxx.lanl.gov/abs/1101.1645}{{\tt
  arXiv:1101.1645}}]. * Temporary entry *.

\bibitem{CMS-PAS-EXO-11-022}
``{CMS-PAS-EXO-11-022}.''
  \url{http://cdsweb.cern.ch/record/1370057/files/EXO-11-022-pas.pdf}.

\bibitem{Kilic:2008ub}
C.~Kilic, S.~Schumann, and M.~Son, {\it {Searching for Multijet Resonances at
  the LHC}},  {\em JHEP} {\bf 0904} (2009) 128,
  [\href{http://xxx.lanl.gov/abs/0810.5542}{{\tt arXiv:0810.5542}}].

\bibitem{Schumann:2011ji}
S.~Schumann, A.~Renaud, and D.~Zerwas, {\it {Hadronically decaying
  color-adjoint scalars at the LHC}},
  \href{http://xxx.lanl.gov/abs/1108.2957}{{\tt arXiv:1108.2957}}.

\bibitem{Bai:2011mr}
Y.~Bai and J.~Shelton, {\it {Composite Octet Searches with Jet Substructure}},
  \href{http://xxx.lanl.gov/abs/1107.3563}{{\tt arXiv:1107.3563}}.

\bibitem{Choi:2008ub}
S.~Choi, M.~Drees, J.~Kalinowski, J.~Kim, E.~Popenda, {\em et.~al.}, {\it
  {Color-Octet Scalars of N=2 Supersymmetry at the LHC}},  {\em Phys.Lett.}
  {\bf B672} (2009) 246--252, [\href{http://xxx.lanl.gov/abs/0812.3586}{{\tt
  arXiv:0812.3586}}].

\bibitem{Kribs:2007ac}
G.~D. Kribs, E.~Poppitz, and N.~Weiner, {\it {Flavor in supersymmetry with an
  extended R-symmetry}},  {\em Phys.Rev.} {\bf D78} (2008) 055010,
  [\href{http://xxx.lanl.gov/abs/0712.2039}{{\tt arXiv:0712.2039}}].

\bibitem{Plehn:2008ae}
T.~Plehn and T.~M. Tait, {\it {Seeking Sgluons}},  {\em J.Phys.G} {\bf G36}
  (2009) 075001, [\href{http://xxx.lanl.gov/abs/0810.3919}{{\tt
  arXiv:0810.3919}}].

\bibitem{Aad:2011yh}
{\bf ATLAS} Collaboration, G.~Aad, {\it {Search for Massive Colored Scalars in
  Four-Jet Final States in sqrt(s)=7 TeV proton-proton collisions with the
  ATLAS Detector}},  \href{http://xxx.lanl.gov/abs/1110.2693}{{\tt
  arXiv:1110.2693}}. Long author list - awaiting processing.

\bibitem{coloron}
{\bf CMS} Collaboration, {\it Search for pair-produced dijet resonances in
  events with four high pt jets in pp collisions at 7 tev},  Tech. Rep.
  EXO-11-016, 2012.

\bibitem{Aad:2011uv}
{\bf ATLAS} Collaboration, G.~Aad {\em et.~al.}, {\it {Search for pair
  production of first or second generation leptoquarks in proton-proton
  collisions at sqrt(s)=7 TeV using the ATLAS detector at the LHC}},  {\em
  Phys.Rev.} {\bf D83} (2011) 112006,
  [\href{http://xxx.lanl.gov/abs/1104.4481}{{\tt arXiv:1104.4481}}]. *
  Temporary entry *.

\bibitem{Khachatryan:2010mp}
{\bf CMS} Collaboration, V.~Khachatryan {\em et.~al.}, {\it {Search for Pair
  Production of First-Generation Scalar Leptoquarks in pp Collisions at sqrt(s)
  = 7 TeV}},  {\em Phys.Rev.Lett.} {\bf 106} (2011) 201802,
  [\href{http://xxx.lanl.gov/abs/1012.4031}{{\tt arXiv:1012.4031}}].

\bibitem{Khachatryan:2010mq}
{\bf CMS} Collaboration, V.~Khachatryan {\em et.~al.}, {\it {Search for Pair
  Production of Second-Generation Scalar Leptoquarks in pp Collisions at
  sqrt(s) = 7 TeV}},  {\em Phys.Rev.Lett.} {\bf 106} (2011) 201803,
  [\href{http://xxx.lanl.gov/abs/1012.4033}{{\tt arXiv:1012.4033}}].

\bibitem{Aad:2012cy}
{\bf ATLAS} Collaboration, G.~Aad {\em et.~al.}, {\it {Search for second
  generation scalar leptoquarks in pp collisions at sqrt(s) = 7 TeV with the
  ATLAS detector}},  \href{http://xxx.lanl.gov/abs/1203.3172}{{\tt
  arXiv:1203.3172}}.

\bibitem{ATLAS-CONF-2012-033}
{\it Search for squarks and gluinos using final states with jets and missing
  transverse momentum with the atlas detector in �s = 7 tev proton-proton
  collisions},  Tech. Rep. ATLAS-CONF-2012-033, CERN, Geneva, Mar, 2012.

\bibitem{Meade:2006dw}
P.~Meade and M.~Reece, {\it {Top partners at the LHC: Spin and mass
  measurement}},  {\em Phys.Rev.} {\bf D74} (2006) 015010,
  [\href{http://xxx.lanl.gov/abs/hep-ph/0601124}{{\tt hep-ph/0601124}}].

\bibitem{Matsumoto:2006ws}
S.~Matsumoto, M.~M. Nojiri, and D.~Nomura, {\it {Hunting for the Top Partner in
  the Littlest Higgs Model with T-parity at the CERN LHC}},  {\em Phys.Rev.}
  {\bf D75} (2007) 055006, [\href{http://xxx.lanl.gov/abs/hep-ph/0612249}{{\tt
  hep-ph/0612249}}].

\bibitem{Kong:2007uu}
K.~Kong and S.~C. Park, {\it {Phenomenology of Top partners at the ILC}},  {\em
  JHEP} {\bf 0708} (2007) 038,
  [\href{http://xxx.lanl.gov/abs/hep-ph/0703057}{{\tt hep-ph/0703057}}].

\bibitem{Han:2008gy}
T.~Han, R.~Mahbubani, D.~G. Walker, and L.-T. Wang, {\it {Top Quark Pair plus
  Large Missing Energy at the LHC}},  {\em JHEP} {\bf 0905} (2009) 117,
  [\href{http://xxx.lanl.gov/abs/0803.3820}{{\tt arXiv:0803.3820}}].

\bibitem{Plehn:2010st}
T.~Plehn, M.~Spannowsky, M.~Takeuchi, and D.~Zerwas, {\it {Stop Reconstruction
  with Tagged Tops}},  {\em JHEP} {\bf 1010} (2010) 078,
  [\href{http://xxx.lanl.gov/abs/1006.2833}{{\tt arXiv:1006.2833}}].

\bibitem{Alwall:2010jc}
J.~Alwall, J.~L. Feng, J.~Kumar, and S.~Su, {\it {Dark Matter-Motivated
  Searches for Exotic 4th Generation Quarks in Tevatron and Early LHC Data}},
  {\em Phys.Rev.} {\bf D81} (2010) 114027,
  [\href{http://xxx.lanl.gov/abs/1002.3366}{{\tt arXiv:1002.3366}}].

\bibitem{Aad:2011cw}
{\bf ATLAS} Collaboration, G.~Aad {\em et.~al.}, {\it {Search for scalar bottom
  pair production with the ATLAS detector in pp Collisions at sqrt{s} = 7
  TeV}},  \href{http://xxx.lanl.gov/abs/1112.3832}{{\tt arXiv:1112.3832}}.

\bibitem{Aad:2011wc}
{\bf ATLAS} Collaboration, G.~Aad {\em et.~al.}, {\it {Search for New Phenomena
  in ttbar Events With Large Missing Transverse Momentum in Proton-Proton
  Collisions at sqrt(s) = 7 TeV with the ATLAS Detector}},
  \href{http://xxx.lanl.gov/abs/1109.4725}{{\tt arXiv:1109.4725}}.

\bibitem{Lillie:2007hd}
B.~Lillie, J.~Shu, and T.~M. Tait, {\it {Top Compositeness at the Tevatron and
  LHC}},  {\em JHEP} {\bf 0804} (2008) 087,
  [\href{http://xxx.lanl.gov/abs/0712.3057}{{\tt arXiv:0712.3057}}].

\bibitem{Gregoire:2008mr}
T.~Gregoire and E.~Katz, {\it {A Composite gluino at the LHC}},  {\em JHEP}
  {\bf 0812} (2008) 084, [\href{http://xxx.lanl.gov/abs/0801.4799}{{\tt
  arXiv:0801.4799}}].

\bibitem{Chen:2008hh}
C.-R. Chen, W.~Klemm, V.~Rentala, and K.~Wang, {\it {Color Sextet Scalars at
  the CERN Large Hadron Collider}},  {\em Phys.Rev.} {\bf D79} (2009) 054002,
  [\href{http://xxx.lanl.gov/abs/0811.2105}{{\tt arXiv:0811.2105}}].

\bibitem{Acharya:2009gb}
B.~S. Acharya, P.~Grajek, G.~L. Kane, E.~Kuflik, K.~Suruliz, {\em et.~al.},
  {\it {Identifying Multi-Top Events from Gluino Decay at the LHC}},
  \href{http://xxx.lanl.gov/abs/0901.3367}{{\tt arXiv:0901.3367}}.

\bibitem{Kumar:2009vs}
K.~Kumar, T.~M. Tait, and R.~Vega-Morales, {\it {Manifestations of Top
  Compositeness at Colliders}},  {\em JHEP} {\bf 0905} (2009) 022,
  [\href{http://xxx.lanl.gov/abs/0901.3808}{{\tt arXiv:0901.3808}}].

\bibitem{Gregoire:2011ka}
T.~Gregoire, E.~Katz, and V.~Sanz, {\it {Toptet}},
  \href{http://xxx.lanl.gov/abs/1101.1294}{{\tt arXiv:1101.1294}}.

\bibitem{Kane:2011zd}
G.~L. Kane, E.~Kuflik, R.~Lu, and L.-T. Wang, {\it {Top Channel for Early SUSY
  Discovery at the LHC}},  {\em Phys.Rev.} {\bf D84} (2011) 095004,
  [\href{http://xxx.lanl.gov/abs/1101.1963}{{\tt arXiv:1101.1963}}].

\bibitem{Essig:2011qg}
R.~Essig, E.~Izaguirre, J.~Kaplan, and J.~G. Wacker, {\it {Heavy Flavor
  Simplified Models at the LHC}},
  \href{http://xxx.lanl.gov/abs/1110.6443}{{\tt arXiv:1110.6443}}.

\bibitem{Brust:2011tb}
C.~Brust, A.~Katz, S.~Lawrence, and R.~Sundrum, {\it {SUSY, the Third
  Generation and the LHC}},  \href{http://xxx.lanl.gov/abs/1110.6670}{{\tt
  arXiv:1110.6670}}.

\bibitem{Berger:2011af}
J.~Berger, M.~Perelstein, M.~Saelim, and A.~Spray, {\it {Boosted Tops from
  Gluino Decays}},  \href{http://xxx.lanl.gov/abs/1111.6594}{{\tt
  arXiv:1111.6594}}.

\bibitem{ATLAS-CONF-2012-003}
{\it Search for supersymmetry in pp collisions at sqr(s)=7 tev in final states
  with missing transverse momentum and b-jets with the atlas detector},  Tech.
  Rep. ATLAS-CONF-2012-003, CERN, Geneva, Feb, 2012.

\bibitem{CMS-PAS-SUS-11-020}
{\bf CMS} Collaboration, {\it Search for new physics in events with same-sign
  dileptons, b-tagged jets and missing energy},  Tech. Rep. CMS-PAS-SUS-11-020,
  CERN, Geneva, Mar, 2012.

\bibitem{CDFtrilepton}
{\bf CDF} Collaboration, ``{Search for trilepton new physics and
  chargino-neutralino production at the Collider Detector at Fermilab}.'' CDF
  Note 10636,
  \url{www-cdf.fnal.gov/physics/exotic/r2a/20110826.trilepton_6fb/cdf10636.pdf}.

\end{thebibliography}\endgroup

\end{document}